\documentclass[onecolumn,floatfix,a4paper,
               showpacs,nofootinbib,preprint]{revtex4}

\usepackage[english]{babel}
\usepackage{amsfonts}
\usepackage{amssymb}
\usepackage{amsmath}
\usepackage{hyperref}
\usepackage{graphicx}
\usepackage{graphics,wrapfig}

\usepackage{color}

\newcommand{\bs}{\boldsymbol}
\newcommand*{\dis}{\displaystyle}
\newcommand{\bvar}{\mathbf}

\begin{document}

\title{
Mean-field approach in the multi-component gas of interacting
particles applied to relativistic heavy-ion collisions }

\author{D. Anchishkin$^{1,2,4}$, V. Vovchenko$^{2,3,4,5}$ }

\affiliation{
$^1$Bogolyubov Institute for Theoretical Physics, 03680 Kiev, Ukraine}
\affiliation{
$^2$Taras Shevchenko Kiev National University, 03022 Kiev, Ukraine}
\affiliation{
$^3$GSI Helmholtzzentrum f\"ur Schwerionenforschung GmbH, D-64291 Darmstadt, Germany}
\affiliation{
$^4$Frankfurt Institute for Advanced Studies, D-60438 Frankfurt, Germany}
\affiliation{
$^5$Johann Wolfgang Goethe University, D-60325 Frankfurt, Germany}

\begin{abstract}
Generalized mean-field approach for thermodynamic description of
relativistic single- and multi-component gas in the grand canonical
ensemble is formulated.
In the framework of the proposed approach different phenomenological
excluded-volume procedures are presented and compared to the existing ones.
The mean-field approach is then used to effectively include hard-core repulsion
in hadron-resonance gas model for description of chemical freeze-out in
heavy-ion collisions.
We calculate the collision energy dependence of several quantities for different
values of hard-core hadron radius and for different excluded-volume procedures
such as the van der Waals and Carnahan-Starling models.
It is shown that a choice of the excluded-volume model becomes important for
large particle densities.
For large enough values of hadron radii
($r\gtrsim0.9$~fm) there can be a sizable difference between different
excluded-volume procedures used to describe the chemical freeze-out in
heavy-ion collisions.
At the same time, for the smaller and more commonly used values of hard-core hadron radii
($r\lesssim0.5$~fm),
the precision of the van der Waals excluded-volume procedure is
shown to be sufficient.
\end{abstract}


\maketitle

\section{Introduction}
\label{sec:introd}
Thermodynamic models of description of properties of the strongly interacting
matter are among the most valuable tools in modern high-energy physics.
The hadron-resonance gas model and its modifications have been successfully used
to extract thermodynamic parameters of matter created in heavy-ion collisions,
by fitting the rich data on mean hadron multiplicities in various experiments
ranging from low energies at SchwerIonen-Synchrotron (SIS) to highest energy of
the Large Hadron Collider
(LHC)~\cite{CleymansSatz,CleymansRedlich1998,CleymansRedlich1999,
Becattini2001,BraunMunzinger2001,Becattini2004,ABS2006,ABS2009}.
Various equations-of-state of the strongly interacting matter (either
phenomenological or based on lattice QCD) are used as an input into fluid
dynamical models, which describe the dynamics of nucleus-nucleus collisions.
In the simplest case, the hadronic phase is described by the multi-component
ideal gas of point-like hadrons.
In more realistic model one needs to take into account the attractive and
repulsive interactions between hadrons.
According to the arguments of Dashen, Ma, and Bernstein~\cite{DMB}, the
inclusion into the model of all known resonances as free particles allows to
effectively model the attraction between hadrons.
In order to describe the repulsive part of hadronic interaction, various
phenomenological excluded-volume procedures have been
proposed~\cite{Hagedorn,CleymansSuhonen,Rischke1991,UddinSingh}.
Another way of modeling the attractive and repulsive interactions is the
relativistic mean-field theory such as Walecka model~\cite{wal,ser} and
its generalizations.
The excluded-volume effects were also treated in the framework of
generalized mean-field approach~\cite{anch,anchsu,TSPS1998,TPS1998}
where the resulting temperature-dependent mean-field allows to perform
an excluded-volume procedure of van der Waals type.
In the present work we extend this approach by formulating the generalized
mean-field theory for single- and multi-component gases of particles from
basic thermodynamic considerations.
We show that the presented approach allows one to conveniently formulate
various models of classical equation of state for real gases in the grand canonical ensemble and in a thermodynamically consistent way.
As an example, we consider
various excluded-volume procedures, such as the van der Waals or Carnahan-Starling
models, in the framework of the hadron-resonance gas model.
The importance of thermodynamic consistency constraint in phenomenological
excluded-volume models has recently been discussed in Ref.~\cite{Gorenstein2012}.

The paper is organized as follows.
In Sec.~\ref{sec2} we formulate the self-consistent mean-field approach from thermodynamic considerations.
In Sec.~\ref{sec3} we connect our approach to the virial expansion of the
interacting classical gas.
In Sec.~\ref{sec4} we generalize the mean-field theory for a multi-component
system.
In Sec.~\ref{sec5} different excluded-volume procedures for systems with
repulsive interaction are formulated in terms of mean fields.
In Sec.~\ref{sec6} we perform calculations in the mean-field version of the
hadron-resonance gas model for the description of chemical freeze-out in
heavy-ion collisions. Sec.~\ref{sec7} closes the article with conclusions.

\section{Self-consistent statistical mechanics in framework of mean-field approach}
\label{sec2}
Let us consider the system of interacting particles from general thermodynamic
point of view.
We start from the {\it canonical ensemble} and our
consideration will be done in terms of the density of free energy
$\phi (n,T)$ (we reserve the letter "f" for the distribution function), which
depends on the density of particles $n$ and temperature $T$.
Description in terms of the density of free energy (DFE)
gives complete information about the many-particle system, in particular
DFE relates to the main thermodynamic quantities in the following way
\begin{eqnarray}
\label{eq:phi-1}
\phi (n,T) &=& \varepsilon(n,T)\, -\, T\, s(n,T)\,,
\\
\phi (n,T) &=& n\, \mu(n,T)\, -\, p(n,T)\,,
\label{eq:phi-2}
\end{eqnarray}
where $\varepsilon(n,T)$ is the energy density, $p(n,T)$ is the pressure.
We adopt the system of units $k_{_B} = c = \hbar = 1$.
Two quantities $\mu(n,T)$ (the chemical potential) and  $s(n,T)$ (the entropy density)
are given as partial derivatives with respect to the independent variables $(n,\, T)$
\begin{eqnarray}
\mu \ =\ \left( \frac{\partial \phi }{\partial n} \right)_T \,, \qquad
s\ =\ -\, \left( \frac{\partial \phi }{\partial T} \right)_n \,.
\label{eq:phi-3}
\end{eqnarray}

For a system of interacting particles the DFE can be written as a sum of free
and interacting contributions
\begin{equation}
\phi (n,T)\ =\ \phi_0(n,T)\, +\, \phi_{\rm int}(n,T)\ ,
\label{eq:phi-4}
\end{equation}
where $\phi_0$ is the DFE of the ideal system (without interaction).
It follows from \eqref{eq:phi-3}, that the chemical potential can also be
split into ``free'' and ``interacting'' pieces
\begin{equation}
\mu \, =\, \mu_0 + \left( \frac{\partial \phi _{\rm int}}{\partial n} \right) _{T} \ ,
\quad {\rm where} \quad
\mu_0\  \equiv \ \left(\frac{\partial \phi_0}{\partial n}\right)_T \ .
\label{eq:mu-int}
\end{equation}
Taking into account Eqs. (\ref{eq:phi-2}), (\ref{eq:phi-4}) and
(\ref{eq:mu-int}) for the system of interacting particles one can represent the
pressure in the following form
\begin{equation}
p\ =\
n\, \mu(n,T)\, -\, \phi(n,T) \
=\
p_0(n,T) + n\, \left( \frac{\partial \phi _{\rm int}}{\partial n} \right)_T
- \phi _{\rm int}\ ,
\label{eq:pressure-int}
\end{equation}
where
\begin{equation}
p_0(n,T)\ =\ n\, \mu_0(n,T)\, -\, \phi_0(n,T) \ .
\label{eq:pressure0-1}
\end{equation}
Here the independent variables $n$ and $T$ correspond to the interacting system.
The ``free'' pressure $p_0$ from the last equation (\ref{eq:pressure0-1}) is the
pressure of the ideal gas calculated for the same values of thermodynamic
variables $n$ and $T$
\begin{equation}
p_0(T,\mu_0)\, =\, \frac{g}{3} \int \frac{d^3k}{(2\pi )^3}\,
\frac{{\bvar k}^2}{\sqrt{{m^2 + \bvar k}^2} }\, f_0[\bvar k;T,\mu_0(n,T)] \ ,
\label{eq:pressure0-2}
\end{equation}
where $g$ is the degeneracy factor, $f_0(\bvar k;T,\mu_0)$ is the ideal gas
distribution function (the Boltzmann, Fermi-Dirac or Bose-Einstein one).

By definition we introduce now the following important notations:
\begin{eqnarray}
\label{eq:mean-field}
U(n,T) & \equiv &
\left[ \frac{\partial \phi_{\rm int}(n,T)}{\partial n} \right]_T \ ,
\\
P^{\rm ex}(n,T) & \equiv &
n \, \left[ \frac{\partial \phi_{\rm int}(n,T)}{\partial n} \right]_T \,
-\, \phi_{\rm int}(n,T)\ .
\label{eq:ex-pressure}
\end{eqnarray}
We immediately fix that these two quantities are related to one another by
equality
\begin{equation}
n \frac{\partial U(n,T)}{\partial n}\ =\ \frac{\partial P^{\rm ex}(n,T)}{\partial n} \ .
\label{eq:consistency}
\end{equation}

Substituting $P^{\rm ex}(n,T)$ from (\ref{eq:ex-pressure}) into
Eq. (\ref{eq:pressure-int}) we obtain
\begin{equation}
p\ =\ p_0(T,\mu_0) + P^{\rm ex}(n,T)\ .
\label{eq:pressure-int-2}
\end{equation}
Evidently, if in this equation one regards $p_0(T,\mu_0)$ as the pressure of the
ideal gas, then, the quantity $P^{\rm ex}(n,T)$ should be treated as the
{\it excess pressure}
which appears
due to presence of interactions.

It is evident from Eqs.~\eqref{eq:mean-field}
and \eqref{eq:ex-pressure} that the knowledge of both $U(n,T)$ and $P^{\rm ex}(n,T)$ is equivalent
to the knowledge of DFE, and therefore mean field and excess pressure contain
full information about the system.
Next, we would like to switch to the
{\it grand canonical ensemble}, which is more convenient for describing
the many-particle systems with a variable number of particles.
In this ensemble the independent variables are temperature $T$
and chemical potential $\mu$. The pressure $p(T,\mu)$ given
in terms of these independent variables contains complete information
about the system.
The basis of the transition from the canonical ensemble to the grand canonical
ensemble is an equivalence of thermodynamic averages in both ensembles.
In this case the particle density becomes a function of
$T$ and $\mu$, i.e. $n=n(T,\mu)$, and the pressure $p(T,\mu)$
is determined from \eqref{eq:pressure-int-2} as
\begin{equation}
p(T,\mu)\, =\, p_0[T,\mu_0(n,T)] + P^{\rm ex}[n(T,\mu),T] \,.
\end{equation}
The ``free'' pressure $p_0$ is determined by (\ref{eq:pressure0-2})
where the ``free'' chemical potential $\mu_0$ can be expressed
in terms of $T$ and $\mu$ by substituting notation (\ref{eq:mean-field}) into
Eq.~(\ref{eq:mu-int})
\begin{equation}
\mu_0(T,\mu)\ =\ \mu \, -\, U[n(T,\mu),T] \ .
\label{eq:mu0-2}
\end{equation}
The pressure of the interacting gas is calculated using
Eqs.~(\ref{eq:pressure-int-2}) and (\ref{eq:pressure0-2}), where we substitute into
(\ref{eq:pressure0-2}) the free chemical potential $\mu_0$ from (\ref{eq:mu0-2}).
As a result, we obtain the following expression for the pressure of the
gas of interacting particles in the grand canonical ensemble
\begin{equation}
p(T,\mu )\ =\
 \frac{g}{3} \int \frac{d^3k}{(2\pi)^3}\, \frac{{\bvar k}^2}
{\sqrt{m^2 + {\bvar k}^2} }\, f(\bvar k;T,\mu,n) + P^{\rm ex}(n,T)\ ,
\label{eq:pressure-int-3}
\end{equation}
where it is implied that $n = \frac{\langle N \rangle}{V} = n(T,\mu)$ and
\begin{equation}
f(\bvar k;T,\mu,n)\, =\,  \left\{ \exp{\left[ \frac{\sqrt{m^2 + {\bvar k}^2} +
       U(n,T)-\mu }{T}\right]} + a \right\}^{-1}
\label{eq:df}
\end{equation}
with $a=+1$ for fermions, $a=-1$ for bosons and $a=0$ for the Boltzmann
approximation.

It is an easy task to go to the nonrelativistic sector, where $|\bvar k| \ll m$.
Indeed, we just make an alteration in the dispersion law
\begin{equation}
\sqrt{m^2 + {\bvar k}^2} \ \to \ m\, +\, \frac{\bvar k^2}{2m} \,,
\label{eq:dispersion-law}
\end{equation}
and take into account that the ``nonrelativistic'' chemical potential,
$\widetilde{\mu}$, relates to ``relativistic'' one, $\mu$, as
$\widetilde{\mu} = \mu-m$.

In a sense, present transition from the canonical ensemble to the grand
canonical ensemble is based on the alteration of the particle number density $N/V$
as an independent variable by its mean value  $\langle N \rangle/V$ which
depends on chemical potential $\mu$ and temperature $T$.
Hence, from now on, under the quantity $n$ we adopt the mean value
$n = \langle N \rangle/V$.
Essentially, this transition corresponds to the Legendre transform of the
thermodynamic potentials, namely from description of the system by means of the
density of free energy $\phi(n,T)$ we switch to the description
by means of
pressure $p(T,\mu)$ or by thermodynamic potential $\Omega = - p(T,\mu) V$.
In general case such a transform means
\begin{equation}
p(T,\mu) \, =\, n\, \mu\, -\, \phi(n,T)\,,
\qquad  {\rm where}  \qquad
\mu \, =\, \left[ \frac{\partial \phi(n,T) }{\partial n} \right]_T \,.
\label{eq:legendre-transform1}
\end{equation}
Actually, for a single-component gas the thermodynamic limit assumption,
i.e. $N/V = $~const when $N \to \infty$ and $V \to \infty$, is
necessary for this transition to be valid.

Because we separate the free energy in two pieces,
$\phi (n,T) = \phi_0(n,T) + \phi_{\rm int}(n,T)$, in our case the Legendre
transform looks like
\begin{equation}
p(T,\mu)\, - p_0(T,\mu_0) = n\, (\mu - \mu_0)\, -\, \phi_{\rm int}(n,T)\ ,
\quad  {\rm where}  \quad
\mu - \mu_0  = \left[ \frac{\partial \phi_{\rm int}(n,T) }{\partial n} \right]_T \,.
\label{eq:legendre-transform2}
\end{equation}
Next, we do not resolve second equation in (\ref{eq:legendre-transform2})
explicitly, but keep it in an implicit form introducing short-hand notations
(\ref{eq:mean-field}) and (\ref{eq:ex-pressure}).
After that Eq.~(\ref{eq:legendre-transform2}) becomes
\begin{equation}
p(T,\mu)\, =\,   p_0[T,\mu - U(n,T)] + P^{\rm ex}(n,T)\ ,
\label{eq:legendre-transform3}
\end{equation}
where particle density carries dependence on temperature and chemical potential
in implicit form, $n(T,\mu)$.
Our next step is to determine the function $U(n,T)$ or the function
$P^{\rm ex}(n,T)$ from another approach.
For instance, one can determine these functions from the virial expansion
(see more details below),
or from some phenomenological equation
of state which describes the real gases in the canonical ensemble.
When one gets the function $U(n,T)$ or the function $P^{\rm ex}(n,T)$ in explicit
form just after that we can say that the problem is formulated in a closed form.

Expression \eqref{eq:pressure-int-3} for pressure contains a particle number
density $n(T,\mu)$, which is a hitherto unknown quantity at given $T$ and $\mu$.
In order to determine $n(T,\mu)$ we recall the following thermodynamic identity
\begin{equation}
n(T,\mu)\ =\ \frac{\partial p(T,\mu)}{\partial \mu} \,.
\label{eq:n-as-deriv}
\end{equation}
To proceed in calculation of (\ref{eq:n-as-deriv}) it is convenient to use
expression of the pressure (\ref{eq:pressure-int-3}) in the equivalent form
\begin{equation}
p(T,\mu)\ =\ \frac{gT}{a} \int \frac{d^3k}{(2\pi)^3}\,
\ln{\left\{ 1\, +\, a \exp{\left[ \frac{-\sqrt{m^2 + {\bs k}^2}
     - U(n,T) + \mu }{T}\right]} \right\}} + P^{\rm ex}(n,T)\ .
\label{eq:p-as-log}
\end{equation}
Taking derivative of this expression with respect to $\mu$ and using relation
\eqref{eq:consistency} one obtains
\begin{equation}
n\ =\ g  \int \frac{d^3k}{(2\pi)^3}\, f(\bvar k;T,\mu,n)\ .
\label{eq:n}
\end{equation}
In this sense we can regard the statistical mechanics under consideration as a
nonlinear one because right-hand-side of expression (\ref{eq:n}) for density $n$
itself depends explicitly on $n$. Therefore, the obtained expression is in fact a
nonlinear equation with respect to $n$.
It is also seen that the function $f(\bvar k;T,\mu,n)$ given in (\ref{eq:df}) and
which appears in \eqref{eq:n} can be regarded as the distribution function with
dispersion relation $\epsilon(\bs k) = \epsilon_0(\bs k) + U(n,T)$, where
$\epsilon_0(\bs k) = \sqrt{m^2+\bs k^2}$.
Consequently, we can interpret $U(n,T)$ as a {\it thermodynamic mean field}
which corrects the free single-particle spectrum $\epsilon_0(\bs k)$ by
accounting for interaction of particles in the many-particle system.

To determine the energy density $\varepsilon$, we use the Euler relation, $\varepsilon + p = Ts + \mu n$,
where $s = \partial p/\partial T$. One obtains
\begin{eqnarray}
\varepsilon(T,\mu ) & = & g \int \frac{d^3k}{(2\pi )^3}\,
\sqrt{m^2 + {\bvar k}^2}\, f(\bs k;T,\mu,n)\, -\, P^{\rm ex}(n,T)\, +\,  n\, U(n,T)\, +
\nonumber \\
& & +\, T\left\{ \, \left[ \frac{\partial P^{\rm ex}(n,T)}{\partial T}\right]_n \,
-\, n\, \left[ \frac{\partial U(n,T)}{\partial T} \right]_n \right\} \ .
\label{eq:energy-den-int}
\end{eqnarray}

The self-consistency of the thermodynamic picture that we obtain above
is due to relation (\ref{eq:consistency}) between the mean field $U(n,T)$ and
the excess pressure $P^{\rm ex}(n,T)$.
That is why it is reasonable to regard relation (\ref{eq:consistency}) as a
{\it condition of thermodynamic consistency}.

On the other hand, one can start to build a model on the base of some
phenomenological approach in which he can determine
$U(n,T)$ or/and $P^{\rm ex}(n,T)$.
Then, relation (\ref{eq:consistency}) needs to be adopted by definition.
Indeed, taking the partial derivative of thermodynamic potential $\Omega(T,\mu, V)$
and using this relation
it is possible to obtain the correct expression for the particle number density, i.e.
$- \left[ \partial \Omega/\partial \mu \right]_{T,V} = nV$.
Meanwhile, in our approach relation (\ref{eq:consistency}) was obtained
without any approximations
as an intrinsic property of the mean field $U(n,T)$ and the excess pressure
$P^{\rm ex}(n,T)$ introduced in (\ref{eq:mean-field}) and (\ref{eq:ex-pressure}).

Attention should be drawn to the fact that for a given function
$\phi_{\rm int}(n,T)$ (as the starting point of the problem) or, which is the
same, for a given mean field $U(n,T)$, Eq.~(\ref{eq:n}) is not a function but a nonl-linear equation for the particle number density $n$.
This equation has to be solved in a self-consistent way for any given point in the
$(T,\mu)$ plane.
The solution will result in the explicit dependence $n=n(T,\mu)$, which in general
differs from the ideal gas dependence, $n_0(T,\mu_0)$.
With known $n(T,\mu)$, the pressure $p(T,\mu)$ and the energy density $\varepsilon(T,\mu )$ can be determined from equations (\ref{eq:pressure-int-3}) and (\ref{eq:energy-den-int}), respectively.
And in the end one can say, that performing the steps listed above we reformulate
a statistical description of the system of interacting particles to a description
by means of a nonlinear statistical approach.

Analogous approach has been of wide use in relativistic mean field theories
\cite{wal,ser,wald} where particles interact by means of a scalar field $\phi $ (attraction)
and a vector field $V_{\mu }$ (repulsion).
Due to the rotation invariance just the ``time'' component of the vector field,
$V_0$, survives.
In our notations it corresponds to the mean field $U(n,T)$.

\section{Single particle nonlinear representation
\newline
\hspace{20mm} of interacting classical gas}
\label{sec3}
To illustrate the above considerations we consider
a classical gas with particle repulsion.
The grand partition function of the classical gas is
\begin{eqnarray}
\Xi (T,\mu ,V) & = & \sum _{N=0}^{\infty }V^{N}g^{N}\, \int
\frac{d^3k_1}{(2\pi )^3} \cdots \frac{d^3k_N}{(2\pi )^3}\,
\exp{\left\{ -\frac{1}{T} \left[  \sum _{j=1}^{N} e_{0}(k_{j})-\mu N
\right] \right\}} \nonumber \\
& & \times\, \frac{1}{V^N N!} \int d^3r_1 \cdots d^3r_N\,
\exp{\left( -\frac{U_N}{T} \right) }\ ,
\label{30a}
\end{eqnarray}
where $e_0(k)={\bvar k}^2/2m$ in the nonrelativistic case and
$e_0(k)=\sqrt{ m^2 + {\bf k}^2}$ in the relativistic one.
Here the potential energy of the $N$-particle system reads as
\begin{equation}
U_N\, =\, \sum _{i<j}^N \,\Phi \left(|{\bvar r}_i-{\bvar r}_j|\right)\ ,
\label{30b}
\end{equation}
where $\Phi (|{\bf r}|)$ is the two-particle potential.

On the next step we use the first and second Mayer's theorems \cite{mayer,kuni}
which are the crucial points of our consideration.
Due to these theorems the virial expansion of the grand partition function
$\Xi(T,\, n)$ for the Boltzmann gas with two-particle interaction reads
\begin{equation}
\ln{ \Xi }\, =\, V\, \left[ n + \sum_{i=2}^{\infty}\, B_i(T)\, n^i\, \right] \ ,
\label{eq:Xi}
\end{equation}
\begin{equation}
\ln{ z_0 }\, =\, \ln{n}\, +\, \sum_{i=2}^{\infty}\, \frac{i}{i-1}\, B_{i}(T)\, n^{i-1} \ ,
\label{31}
\end{equation}
where $z_{0}$ is the ideal single-particle partition function
\begin{equation}
z_0(T,\mu)\ =\ g\int  \frac{d^{3}k}{(2\pi )^3}\,
\exp{ \left[-\frac{e_0(\bs k) - \mu }{T} \right] }\ .
\label{32}
\end{equation}
The virial coefficients $B_{i}(T)$  depend on temperature $T$ , e.g.,
the second virial coefficient for particles interacting through the potential
$\Phi (r)$  is given by
\begin{equation}
B_{2}(T)\, =\, \frac{1}{2} \int d^3r\,
\left\{ 1-\exp{\left[-\frac{\Phi (r)}{T} \right]} \right\} \,.
\label{33}
\end{equation}
The expansions entering on the r.h.s. of Eqs.~(\ref{eq:Xi}) and (\ref{31}),
can be incorporated into the quantities
\begin{equation}
P_{\rm cl}^{\rm ex}(n,T)\, =\, T\, \sum^{\infty }_{i=2}\, B_i(T)\, n^i \ ,
\label{eq:Pcl-def}
\end{equation}
\begin{equation}
U_{\rm cl}(n,T)\, =\, T\, \sum^{\infty }_{i=2}\, \frac{i}{i-1}\, B_i(T)\, n^{i-1} \ ,
\label{eq:Ucl-def}
\end{equation}
which obey a relation analogous to (\ref{eq:consistency}),
\begin{equation}
n \frac{\partial U_{cl}(n,T)}{\partial n}\
=\ \frac{\partial P_{\rm cl}^{\rm ex}(n,T)}{\partial n} \ .
\label{eq:PUcl}
\end{equation}
Equation ({\ref{eq:PUcl}}) is valid for every pair of the correspondent terms in
expansions (\ref{eq:Pcl-def}) and (\ref{eq:Ucl-def}) separately. Thus it will still be
valid if one truncates the series at any order.
With the use of these notations, Eqs.~(\ref{eq:Xi}) and (\ref{31}) can be
rewritten as
\begin{equation}
p(T,\mu)\, =\, \frac TV \, \ln{ \Xi }\, =\, T n(T,\mu )\, +\, P_{\rm cl}^{\rm ex}(n,T)\ ,
\label{eq:pcl}
\end{equation}
\begin{equation}
n(T,\mu)\, =\, g\int \frac{d^{3}k}{(2\pi )^3} \exp{\left[ -\frac{e_0(k) - \mu
+ U_{\rm cl}(n,T)}{T} \right]} \ .
\label{eq:ncl}
\end{equation}
These expressions, which determine the thermodynamic behaviour of the
classical (Boltzmann) gas, have been obtained here without any approximations
and can be regarded as the single particle representation of the original
partition function (\ref{30a}).
On the other hand, in view of the last two expressions, one can regard
the quantities $U(n,T)$ and $P_{\rm cl}^{\rm ex}(n,T)$ as ``mean'' field and excess
pressure respectively, but also as rigorous quantities in the framework of
classical statistics. And in addition, if one describes
many-particle system by expressions (\ref{eq:pcl}) and
(\ref{eq:ncl}) then the thermodynamic quantities $U(n,T)$ and $P_{\rm cl}^{\rm ex}(n,T)$
possess the virial expansions (\ref{eq:Pcl-def}) and (\ref{eq:Ucl-def}).

For the energy density $\varepsilon$, using the Euler relation,
$\varepsilon + p = Ts + \mu n$, and $s = \partial p/\partial T$ we obtain
\begin{eqnarray}
\varepsilon (T,\mu ) & = & g \int \frac{d^3k}{(2\pi )^3}\,
e_{0}(k)\, f(\bvar k;T,\mu,n)\, +\,  n\, U_{cl}(n,T)\, -\, P_{\rm cl}^{\rm ex}(n,T)\,  +
\nonumber \\
& & +\, T\left\{ \, \left[ \frac{\partial P_{\rm cl}^{\rm ex}(n,T)}{\partial T}\right]_n \,
-\,  n\, \left[ \frac{\partial U_{\rm cl}(n,T)}{\partial T} \right]_n\right\} \ .
\label{eq:encl}
\end{eqnarray}
It is evidently seen that this expression for the energy density coincide with
Eq.~(\ref{eq:energy-den-int}) obtained in previous section.

\subsection{Correspondence to thermodynamic mean-field theory}

We adjust now the correspondence between the approach which was elaborated
in Sec.~\ref{sec2} and the present one.
For this purpose we consider the first approach in that region where the
classical statistics can be used.
Then, distribution function (\ref{eq:df}) reduces to
\begin{equation}
f(\bvar k; T,\mu ,n)\, =\,
\exp{ \left[ -\frac{e_{0}(k) + U(n,T)-\mu }{T}\right]} \ .
\label{38a}
\end{equation}
We now compare two descriptions for the same system: The first one is based  on
the grand partition function (\ref{30a}), in the form of Eqs.~(\ref{eq:pcl}),
(\ref{eq:ncl}), while the second one uses the distribution function (\ref{38a}) to
calculate pressure (\ref{eq:pressure-int-3}) and particle density (\ref{eq:n}).
If these two approaches give indeed the same result then we obtain
\begin{equation}
U(n,T)=U_{\rm cl}(n,T) \ ,
\label{38b}
\end{equation}
and consequently
\begin{equation}
P^{\rm ex}(n,T)=P_{\rm cl}^{\rm ex}(n,T) \ .
\label{38c}
\end{equation}
This conclusion gives a constructive algorithm to develop a model of the
interacting system by means of a mean field.
Indeed, we can extract the mean field from the known classical equation of state
$p = p_{\rm cl}(n,T)$.
First, the excess pressure is determined as $P^{\rm ex}(n,T) = p_{\rm cl}(n,T) - nT$.
Then, the mean field can be determined from \eqref{eq:consistency} as
\begin{equation}
U(n,T)\ =\ U(n=0,T)\,
+\, \int_0^n \frac{1}{n'}\, \frac{\partial P^{\rm ex}(n',T)}{\partial n'} \, dn' \,.
\label{eq:determ-u}
\end{equation}
Equation \eqref{eq:determ-u} contains the unknown quantity $U(n=0,T)$, which
is an integration constant.
To fix this quantity we recall that systems with short-range interaction should
approach the ideal gas in the low density limit ($n \to 0$).
Since the mean field is zero for ideal gas, then one should fix this constant
as $U(n=0,T) = 0$.

\section{Mean-field approach for multi-component gas}
\label{sec4}
The proposed phenomenological mean-field approach can be generalized for the
case of a mixture of $f$ different particle species.
Let us denote $\bvar n = (n_1, \ldots, n_f)$ as a set of particle densities and
$\bs \mu = (\mu_1, \ldots, \mu_f)$ as a set of chemical potentials corresponding
to each of the particle species.
In a case of a mixture of $f$ particle species there will be $f$ different mean
fields $U_i (\bvar n,T)$ corresponding to each particle species and a common
excess pressure $P^{\rm ex} (\bvar n, T)$.
It is natural that $U_i (\bvar n, T)$ and $P^{\rm ex} (\bvar n, T)$ can depend
on any of the particle densities of different species.
The density of free energy (DFE) in this case reads
\begin{equation}
\phi(\bvar n,T)\ =\ \sum_{i=1}^f n_i \mu_i(\bvar n,T)\, -\, p(\bvar n,T)\,.
\label{eq:dfe-multi}
\end{equation}
For the system of interacting particles the DFE can be expressed as a sum of
free and interacting parts
\begin{equation}
\phi(\bvar n,T)\ =\ \sum_{i=1}^f \phi_0^i(n_i,T)\, +\, \phi_{\rm int}(\bvar n,T)\,.
\label{eq:dfe-interact}
\end{equation}
Using the same considerations as in Section~\ref{sec2} one can switch to the grand
canonical ensemble by putting a mean value of particle number density
$\langle N_i \rangle/V$ in place of $n_i$
in Eqs.~(\ref{eq:dfe-multi}) and (\ref{eq:dfe-interact}) for each of the particle
species ($i=1,\ldots,f$).
In this case the total pressure $p(T, \bs \mu)$ and density $n_i(T, \bs \mu)$ of
particle species $i$ can be expressed as
\begin{eqnarray}
p(T,\bs \mu ) & = &
 \sum_{i=1}^f \, \frac{g_i}{3} \int \frac{d^3k}{(2\pi)^3}\, \frac{{\bvar k}^2}
{\sqrt{m_i^2 + {\bvar k}^2} }\, f_i(\bvar k;T,\bs \mu,\bvar n)\, +\, P^{\rm ex}(\bs n,T)\ ,
\label{eq:pressure-multi}
\\
n_i(T,\bs \mu ) & = & g_i  \int \frac{d^3k}{(2\pi)^3}\, f_i(\bvar k;T,\bs \mu,\bvar n)\,,
\label{eq:densities-multi}
\end{eqnarray}
where $f_i(\bvar k;T,\bs \mu,\bvar n)$ is the distribution function for particle species $i$
\begin{equation}
f_i(\bvar k;T,\bs \mu,\bvar n)\, =\,  \left\{ \exp{\left[ \frac{\sqrt{m_i^2 + {\bs k}^2} +
       U_i[\bs n(T,\bs \mu) ,T]-\mu_i }{T}\right]} + a_i \right\}^{-1}\,.
\label{eq:dfsyst}
\end{equation}
Actually, this means that from now on under the quantity $n_i$ we take the
mean value $n_i = \langle N_i \rangle/V$.
The excess pressure $P^{\rm ex}(\bs n,T)$ and mean fields $U_i(\bs n,T)$
are defined as
\begin{eqnarray}
P^{\rm ex}(\bs n,T) & = & \sum_{i=1}^f n_i \left[\frac{\partial
\phi_{\rm int}(\bs n,T)}{\partial n_i}\right]_{T,n_{j \neq i}}
-\, \phi_{\rm int}(\bs n,T), \\
U_i(\bs n,T) & = & \left[\frac{\partial \phi_{\rm int}(\bs n,T)}
{\partial n_i}\right]_{T,n_{j \neq i}} \,,
\end{eqnarray}
and related to each other through a set of equations corresponding
to thermodynamic consistency:
\begin{equation}
\sum_{j=1}^f \, n_j\, \frac{\partial U_j}{\partial n_i}\,
=\, \frac{\partial P^{\rm ex}}{\partial n_i}, \quad i=1\ldots f.
\label{eq:consistency-multi}
\end{equation}
From the relation $\displaystyle
\frac{\partial^2 P^{\rm ex}}{\partial n_i \partial n_k}
= \frac{\partial^2 P^{\rm ex}}{\partial n_k \partial n_i}$ one can see that
mean fields $U_i$ are not independent but are related to each other via
\begin{equation}
\frac{\partial U_i}{\partial n_j}\ =\ \frac{\partial U_j}{\partial n_i}, \quad i,j=1\ldots f \,.
\label{eq:symmetry}
\end{equation}
This relation can be useful when defining mean fields for multi-component gases
from phenomenological considerations (see also \cite{Biro2003}).

Equation~\eqref{eq:densities-multi} represents system of $f$ equations
(usually transcendental ones) for particle densities $n_i$ which can be solved numerically.
Energy density can be evaluated from thermodynamical relation
$\varepsilon = T s + \sum_{i=1}^{f} \mu_i n_i - p$.
One obtains
\begin{eqnarray}
\varepsilon(T, \bs \mu) & =&  \sum_{i=1}^f g_i \int \frac{d^3k}{(2\pi )^3}\,
\sqrt{m_i^2 + {\bvar k}^2}\, f_i(\bvar k;T,\bs \mu,\bs n)\,
-\, P^{\rm ex}(\bs n,T)\, +\, \sum_{i=1}^f\, n_i \, U_i(\bs n,T)\, +
\nonumber \\
& & \quad + \, T\, \left[ \frac{\partial P^{\rm ex}(\bs n,T)}{\partial T}\,
-\, \sum_{i=1}^f \, n_i \,\frac{\partial U_i(\bs n,T)}{\partial T}\right]\,.
\label{eq:eUq}
\end{eqnarray}

\subsection{Conserved charges}
In relativistic mechanics it is commonly the case that number of particles in
closed system can change, for instance due to creation of pairs of particles
and anti-particles, and instead of particle number conservation there are
conservation laws for the charges.
As an example it can be baryon charge, electric charge and strangeness in
strongly interacting systems.
Consequently, the grand canonical treatment of relativistic many particle system,
which contains several  species of particles, is usually formulated in terms of
independent chemical potentials which correspond to conserved charges
rather than particle numbers.
Let us have $s$ independent conserved charges and let
$\bs{\tilde{\mu}} = (\tilde{\mu}_1,\ldots,\tilde{\mu}_s)$ be corresponding
independent chemical potentials.

If we denote the $j$-th charge of the $i$-th particle species as $Q^j_i$ then
chemical potential $\mu_i$ of $i$-th particle species can be expressed as
\begin{equation}
\mu_i (\bs{\tilde{\mu}})\, =\, \sum_{j=1}^s Q^j_i \tilde{\mu}_j\,.
\label{eq:mumu}
\end{equation}
Density $\rho_j$ of $j$-th charge can be expressed via a set of particle
densities $\{n_i\}$ as
\begin{equation}
\rho_j (T, \bs{\tilde{\mu}})\, =\, \sum_{i=1}^f Q^j_i \, n_i [T, \bs \mu(\bs{\tilde{\mu}})]\,.
\end{equation}
There will be a total of $f$ mean fields $U_i(T,\bs{\tilde{\mu}})$, one for each
of the particle species, also in the case of just $s<f$ independent
chemical potentials.
Let us show that we can satisfy the conditions of thermodynamic
consistency, $\rho_j = \partial p(T, \bs{\tilde{\mu}}) / \partial \tilde{\mu}_j$,
provided that the excess pressure and mean fields satisfy~\eqref{eq:consistency-multi}
and that pressure and particles densities are given by
Eqs.~\eqref{eq:pressure-multi} and \eqref{eq:densities-multi}, but
where $\bs \mu$ depends on $\bs{\tilde{\mu}}$ via \eqref{eq:mumu}.
Indeed, taking pressure (\ref{eq:pressure-multi}) in a ``log'' form
\begin{equation}
p(T,\bs{\tilde{\mu}}) = \sum_{i=1}^f \frac{g_i T}{a_i} \int \frac{d^3k}{(2\pi)^3}
\ln{\left\{ 1 + a_i \exp{\left[ \frac{-\sqrt{m_i^2 + {\bvar k}^2}
     - U_i(\bs n,T) + \mu_i }{T}\right]} \right\}} + P^{\rm ex}(\bs n,T) \,,
\label{eq:p-as-log-multi}
\end{equation}
where $n_i [T, \bs \mu(\bs{\tilde{\mu}})]$, we can calculate derivative
$\partial p(T, \bs{\tilde{\mu}}) / \partial \tilde{\mu}_j$ as
\begin{eqnarray}
\frac{\partial p(T, \bs{\tilde{\mu}})}{\partial \tilde{\mu_j}}
&=& \sum_{i=1}^f
g_i  \int \frac{d^3k}{(2\pi)^3}\, f_i(\bvar k;T,\bs \mu, \bs n)
\left[ Q^j_i - \sum_{k=1}^f \frac{\partial U_i(\bvar n,T)}{\partial n_k}\,
\frac{\partial n_k(T,\bs{\tilde{\mu}})}{\partial \tilde{\mu_j}} \right] \, +
\nonumber\\
& & \qquad
+ \sum_{k=1}^f \frac{\partial P^{\rm ex}(\bvar n,T)}{\partial n_k}\,
\frac{\partial n_k(T,\bs{\tilde{\mu}})}{\partial \tilde{\mu_j}}
\nonumber \\
&=&
\sum_{i=1}^f Q^j_i n_i \, -\, \sum_{k=1}^f \left[ \sum_{i=1}^f n_i\,
\frac{\partial U_i(\bvar n,T)}{\partial n_k}\,
-\, \frac{\partial P^{\rm ex}(\bvar n,T)}{\partial n_k}  \right]\,
\frac{\partial n_k(T,\mu)}{\partial \tilde{\mu_j}} \ ,
\label{eq:deriv-pressure-multi}
\end{eqnarray}
where we have used (\ref{eq:mumu}) which gives $\partial {\mu_i} / \partial {\tilde{\mu}_j} = Q^j_i$.
Taking into account Eq.~\eqref{eq:consistency-multi} we have
\begin{equation}
\frac{\partial p(T, \bs{\tilde{\mu}})}{\partial \tilde{\mu}_j} = \rho_j (T, \bs{\tilde{\mu}}),
\end{equation}
in accordance with thermodynamic relations.
This explains how a
thermodynamic mean-field description of the multi-component systems given by
Eqs.~\eqref{eq:pressure-multi}-\eqref{eq:dfsyst} can be exploited in case of
conserved charges.
%

\section{Excluded-volume procedure}
\label{sec5}
The excluded-volume procedure is a natural way to model
the short-range hard-core repulsion.
In particular, this procedure has been included in different relativistic mean-field models of nuclear matter in order to account for the finite eigenvolume of nucleons, in addition to the description of their interaction through the exchange of mesonic fields (see, e.g., Refs.~\cite{Rischke1991,anchsu,Hempel,Steinheimer2010}).
The thermodynamic mean-field formulation proposed in the present paper
is essentially a microscopic approach, hence it makes the inclusion
of excluded-volume procedure into different microscopic equations of
state especially convenient. 
In the present work, however, we only consider systems with just the hard-core interaction, which is taken into account by means of excluded-volume procedure.

In this section we describe a
way to effectively formulate a description of systems with short-range repulsion as an excluded-volume procedure in the framework of the thermodynamic mean-field model.

\subsection{Mean fields which are proportional to temperature}
Let us consider firstly a single-component classical gas.
In classical systems where repulsive interactions are modeled as
a hard-core repulsion, the virial coefficients $B_i$ are independent of
temperature, see e.g.~\eqref{33}.
It follows from \eqref{eq:Ucl-def} and \eqref{eq:Pcl-def} that, in this case,
the mean field and the excess pressure describing such
a system are linear functions of temperature, i.e.
\begin{equation}
U(n,T)\ \propto \ T \,, \quad \quad \quad P^{\rm ex}(n,T)\, \propto \ T \,,
\label{eq:special-u}
\end{equation}
More generally, let us consider a class of mean fields which are linear in $T$.
One can see from (\ref{eq:energy-den-int}) that in this case the energy density
is determined just by the integral term
\begin{eqnarray}
\varepsilon(T,\mu)\,
=\, g \int \frac{d^3k}{(2\pi)^3}\, \sqrt{m^2 + {\bvar k}^2}\, f(\bvar k;T,\mu,n) \,.
\label{eq:energy-dens-2}
\end{eqnarray}
In case of the Boltzmann statistics for this class of mean fields we find
\begin{eqnarray}
\label{eq:particle-dens-3}
n(T,\mu) &=& e^{-\widetilde{U}(n)}\ n_0(T,\mu)\,,
\\
\varepsilon(T,\mu) &=& e^{-\widetilde{U}(n)}\ \varepsilon_0(T,\mu)\,,
\label{eq:energy-dens-4}
\end{eqnarray}
where $\widetilde{U}(n) = U(T,n)/T$ in accordance with (\ref{eq:special-u})
and we introduce notations
\begin{eqnarray}
\label{eq:particle-dens-5}
n_0(T,\mu)\, & = & \, g \int \frac{d^3k}{(2\pi)^3}\, f_0(\bvar k;T,\mu) \,,
\\
\varepsilon_0(T,\mu)\,
& = &\, g \int \frac{d^3k}{(2\pi)^3}\, \sqrt{m^2 + {\bvar k}^2}\, f_0(\bvar k;T,\mu) \,.
\label{eq:enrgy-dens-6}
\end{eqnarray}
Here $f_0(\bvar k;T,\mu) = \exp{\left[ \frac{-\sqrt{m^2 + {\bvar k}^2} + \mu }{T}\right]}$
is the distribution function of ideal gas at temperature $T$ and chemical potential $\mu$.

One can interpret the factor
\begin{equation}
e^{-\widetilde{U}(n)}\ \equiv \ \vartheta(n)\,,
\label{eq:def-alpha}
\end{equation}
on the r.h.s. of equations (\ref{eq:particle-dens-3}) and (\ref{eq:energy-dens-4})
as a suppression-factor of the free volume $V$ of the system.
Indeed, if we introduce notation of the available volume
\begin{equation}
\widetilde{V}(n) = \vartheta(n)\, V\,,
\label{eq:eff-volume}
\end{equation}
we can rewrite Eqs.~(\ref{eq:particle-dens-3}), (\ref{eq:energy-dens-4})
in the form
\begin{eqnarray}
\label{eq:energy-dens-3a}
N(T,\mu) &=& \widetilde{V}\, n_0(T,\mu)\,,
\\
E(T,\mu) &=& \widetilde{V}\, \varepsilon_0(T,\mu)\,.
\label{eq:energy-dens-4a}
\end{eqnarray}
Obviously, in case of a repulsive interaction the mean field $U(T,n)$  is
positive one and we have
\begin{equation}
0< \vartheta(n) \le 1  \qquad \Rightarrow \qquad
\widetilde{V}\, \le \, V\,.
\label{eq:ineq-volume}
\end{equation}
This means that a repulsive mean field generates an {\it effective} proper
volume $\widetilde{v}_0$ of every particle, which belongs to the system, resulting in the
reduction of the total volume $V$ of the system of these $N$ particles, i.e.
$V \to V-N\widetilde{v}_0 = \widetilde{V}$.
One can then evaluate the effective proper
volume $\widetilde{v}_0(n)$ as
\begin{equation}
\widetilde{v}_0(n) \ =\ \frac 1N \, \left(\,V - \widetilde{V}\, \right) \ =\
v \, \big[\,1 - \vartheta(n)\, \big] \,.
\label{eq:v0}
\end{equation}
where $v = V/N$ is the mean classical volume per particle.
The interpretation of $\widetilde{v}_0$ in accordance with (\ref{eq:v0}) is evident enough.
Indeed, the quantity $V - \widetilde{V}$ represents the total {\it effective} self-volume
of $N$ particles of the system which is due to repulsive interaction between
them.
Then, the {\it effective} single particle self-volume is in one-to-one
correspondence with the original repulsive mean field
\begin{equation}
\widetilde{v}_0 (n) \, =\,  \frac{\dis 1}{\dis n} \, \left\{\,1 - \exp{\left[\,
- \widetilde{U}(n)\, \right]} \, \right\} \,,
\label{eq:v0-2}
\end{equation}
where the r.h.s. of equation depends on the particle density only.

We also note that a presence of a such mean field does not change the average
energy per particle in comparison to ideal gas:
\begin{eqnarray}
\frac{\varepsilon}{n} = \frac{\varepsilon_0}{n_0} = 3T + m \frac{K_1(m/T)}{K_2(m/T)}.
\label{eq:eav}
\end{eqnarray}
So, in some sense, we prove a {\it theorem} which says: \\
If the mean field
is the repulsive one ($U(n,T) > 0$) and it is proportional to the
temperature, i.e. $U(n,T) \propto T$, then, for the Boltzmann
statistics, a presence of the interaction in the system
appears just as reduction of the total volume of the system in the following way:
$\widetilde{V} \to \vartheta(n)\, V$, where $\vartheta(n) = \exp{[-U(n,T)/T]}$.\\
With regard to this, we propose to identify all models where a repulsive mean
field is proportional to temperature as a {\it set of excluded-volume models}.
In this type of models the number of particles and total energy are calculated
by using the ideal gas expressions in the grand canonical ensemble as follows:
\begin{equation}
N(T,\mu)\ =\ \widetilde{V}\, n_0(T,\mu)\,, \quad \quad
E(T,\mu)\ =\ \widetilde{V}\, \varepsilon_0(T,\mu) \,.
\label{eq:N+E}
\end{equation}

In a case of quantum statistics we have
\begin{eqnarray}
\label{eq:energy-dens-7}
n(T,\mu) &=& \vartheta(n) \ \widetilde{n}_0(T,\mu)\,,
\\
\varepsilon(T,\mu) &=& \vartheta(n) \ \widetilde{\varepsilon}_0(T,\mu)\,,
\label{eq:energy-dens-8}
\end{eqnarray}
where
\begin{eqnarray}
\label{eq:energy-dens-9}
\widetilde{n}_0(T,\mu)\, & = &\, g \int \frac{d^3k}{(2\pi)^3}\, \widetilde{f}_0(\bvar k;T,\mu) \,,
\\
\widetilde{\varepsilon}_0(T,\mu)\,
&=&\, g \int \frac{d^3k}{(2\pi)^3}\, \sqrt{m^2 + {\bvar k}^2}\, \widetilde{f}_0(\bvar k;T,\mu) \,.
\label{eq:enrgy-dens-10}
\end{eqnarray}
Here $\widetilde{f}_0(\bvar k;T,\mu)$ is the
distribution function of the Fermi or Bose gas
corrected by a factor $\vartheta(n)$
\begin{equation}
\widetilde{f}_0(\bvar k;T,\mu)\
=\ \left[\dis \exp{\left( \frac{\sqrt{m^2 + {\bvar k}^2}
- \mu }{T}\right)}\, +\, a\, \vartheta(n) \right]^{-1}\,.
\label{eq:f0-quant}
\end{equation}
It turns out that, in addition to volume suppression, here we also effectively
obtain an intermediate quantum statistics because the second term in the
denominator is in the range $0 < \vartheta(n) \le 1$.
We remind that $\vartheta(n) = e^{-\widetilde{U}(n)}$, and $a=1$ for fermions
or $a=-1$ for bosons.
It is interesting to note that in this case of repulsive mean field, when one
increases the particle density $n$ the distribution function (\ref{eq:f0-quant})
shifts closer to the classical Boltzmann one.
Hence, in case of the quantum statistics we can conclude that the repulsive mean
fields which are linear in $T$ reduce the total volume of the system
($V \to \widetilde{V}$), and effectively result in reduction of the effects
connected to quantum statistics.

\subsection{Excluded-volume models for single-component gas}
Let us now consider some particular examples of excluded-volume models
in the framework of the mean-field approach.
As a first example we consider a single-component gas where every particle has
a proper volume $v_0$ which is independent of particle-number density $n$,
i.e. it is some parameter of the model.
Then, the free available volume in the system of $N$ particles is $\widetilde{V} = V - Nv_0$.
Actually, solving (\ref{eq:v0-2}) with respect to mean field $U$ one obtains
\begin{equation}
U(n,T)\ =\ -T \ln{\left( 1-\widetilde{v}_0\, n \right)} \,.
\label{eq:u-dev}
\end{equation}
If here we put approximately $\widetilde{v}_0 \approx v_0 = \,$const, then we come to the
mean-field model which we name as {\it model of directly excluded-volume}, where
$U \equiv U_{\rm de}(n,T) = -T \ln{\left( 1-v_0\, n \right)} $.
In a case of classical gas the equation (\ref{eq:n}) for
particle-number density $n \equiv n_{\rm de}$ can be solved explicitly
\begin{equation}
n_{\rm de}(T,\mu)\ =\ \frac{n^0(T,\mu)}{1 + v_0\,n^0(T,\mu)} \ ,
\label{eq:density-dev}
\end{equation}
where $n^0=z_0$ is the density of point-like particles (Eq.~(\ref{eq:n}) with $U=0$).
This formula demonstrates explicitly our general statement:
switching on the repulsive interaction results in a decrease of the
particle-number density $n_{\rm de}(T,\mu)$ compared to the
particle-number density $n^0(T,\mu)$ of
the ideal gas at the same temperature and chemical potential.
This phenomenon is valid 
in quantum statistics as well.

For the pressure in the Boltzmann sector we obtain using Eq.~(\ref{eq:pcl})
\begin{equation}
p_{\rm de} (T,\mu)\ =\ -\frac{T}{v_0} \ln \left[ 1-v_0\,n_{\rm de}(T,\mu) \right]\
=\ \frac{T}{v_0} \ln \left[ 1 + v_0\,n^0(T,\mu)\right] \ ,
\label{eq:pres-dev}
\end{equation}
where the last equality appears when one substitute $n_{\rm de}(T,\mu)$ from
(\ref{eq:density-dev}).
The energy density in this model is given by
\begin{equation}
\varepsilon_{\rm de}(T,\mu)\  =\ \frac{\varepsilon^0(T,\mu)}{1 + v_0\, n^0(T,\mu)} \ ,
\label{eq:en-density-dev}
\end{equation}
where $\varepsilon^0(T,\mu)$ is the energy density of the Boltzmann gas of
point-like particles.
It is seen that due to hard-core repulsion, which is accounted for by means of
excluded volume, the energy density $\varepsilon_{\rm de}(T,\mu)$ is smaller than
$\varepsilon^0(T,\mu)$ in the same way as a decrease of particle-number density.

It turns out
that pressure (\ref{eq:pres-dev}) of the directly excluded volume
is of the same form as the one in the lattice gas model
\cite{kittel} where the lattice has been constructed from the cells with the
volume $v_0$ and there are $N_{\rm cell}=V/v_0$ cells in the lattice.
The pressure and energy density in this model are due to the
number of possible combinations to place $N$ particles in $N_{\rm cell}$ cells
(configuration entropy).

Next, we consider the 2nd order virial expansion for classical gas of hard spheres of radius~$r$
\begin{equation}
p = T n + T v_0 n^2,
\label{eq:vir}
\end{equation}
where virial coefficient $v_0 = 4\,(4\pi)\,r^3/3$ is four times the intrinsic
volume of a particle.
The excess pressure and mean field are then given by
\begin{eqnarray}
P^{\rm ex}(n,T) & = & T v_0 n^2\,,
\label{eq:Pvir} \\
U(n,T) & = & 2 \, T  v_0  n\,.
\label{eq:Uvir}
\end{eqnarray}
Equations~\eqref{eq:Pvir} and \eqref{eq:Uvir} are general expressions for a second order
virial expansion with virial coefficient $v_0$.
We note that the model of directly excluded-volume is consistent with the 2nd order
virial expansion~\eqref{eq:vir}.
Indeed, if in directly excluded-volume approach one expands the mean field to the
first order with respect to $v_0 n$ and the excess pressure to the second order and lets
$\widetilde{v}_0 = 2 v_0$, then he comes to Eqs.~\eqref{eq:Pvir} and \eqref{eq:Uvir}.

Perhaps the most conventional example of a system with excluded volume is
{\it the van der Waals equation of state}
\begin{equation}
p_{_{\rm VdW}}\ =\ \frac{Tn}{1 - v_0\, n} \ ,
\label{eq:VDW}
\end{equation}
where attractive interactions have been neglected.
Equation~\eqref{eq:VDW} represents extrapolation of Eq.~\eqref{eq:vir} to higher
values of $v_0 n$.
The excess pressure ($P^{\rm ex}=p-nT$) is now
\begin{equation}
P_{\rm VdW}^{\rm ex} (n,T)\ =\ Tn\, \frac{v_0\, n}{1 - v_0\, n} \ .
\label{eq:PVDW}
\end{equation}
Using Eq.~(\ref{eq:determ-u}) with the integration constant $U(n=0,T) = 0$
we obtain the repulsive mean field
\begin{equation}
U_{\rm VdW}(n,T)\ =\ T \frac{v_0\, n}{1-v_0\, n} - T \ln{(1-v_0\, n)} \ .
\label{eq:UVDW}
\end{equation}
For the Boltzmann statistics the particle-number density can be calculated from
\eqref{eq:n} and we get
\begin{eqnarray}
n_{_{\rm VdW}}\ =\ (1\, -\, v_0\, n_{_{\rm VdW}}) \, n^0 (T, \mu)\,
\exp\left(-\frac{v_0 n_{_{\rm VdW}}}{1\, -\, v_0\, n_{_{\rm VdW}}}\right)\,.
\label{eq:nVDW}
\end{eqnarray}
Here $n^0(T,\mu)$ is the particle-number density of a point-like ideal gas and
henceforth the zero superscript will denote the ideal gas quantities.
Then, to obtain particle density $n_{_{\rm VdW}}$ explicitly it is necessary to
solve the transcendental, i.e. nonlinear, equation \eqref{eq:nVDW} with respect
to $n_{_{\rm VdW}}$ for every point $(T,\mu)$

A thermodynamically consistent procedure to incorporate the excluded-volume
correction in the hadron gas models was formulated in Ref. \cite{Rischke1991}
by means of direct calculation of the grand partition function in the Boltzmann
approximation.
The pressure in that approach is given by
\begin{eqnarray}
P^{\rm excl} & = & n^0(T, \mu) \, T \, e^{-v_0 P^{\rm excl} / T},
\label{eq:Pexcl}
\end{eqnarray}
It is the transcendental equation with respect to pressure, $P^{\rm excl}$.
Performing some simple algebra one can easily show
that Eq.~\eqref{eq:Pexcl} can be obtained from Eqs.~\eqref{eq:VDW}
and \eqref{eq:nVDW} and therefore our mean-field approach coincides with
excluded volume procedure from Ref.~\cite{Rischke1991} in the Boltzmann limit.

The last example of an excluded-volume procedure for a single-component gas that
we consider is an approach based on a well-known {\it Carnahan-Starling equation
of state}~\cite{CarnahanStarling}, which goes beyond the van der Waals one and
is known to describe the fluid phase of the hard sphere model more accurately.
The pressure is given in terms of the packing fraction $\eta$ by the
Carnahan-Starling formula
\begin{equation}
P_{\rm CS}\, =\, Tn\, \frac{1 + \eta + \eta^2 - \eta^3}{(1-\eta)^3}\,.
\end{equation}
Packing fraction $\eta$ can be expressed in terms of
second virial coefficient $v_0$ and particle density $n$ as $\eta = v_0 n / 4$.
From here the excess pressure reads as
\begin{equation}
P_{\rm CS}^{\rm ex} (n,T)\ =\ Tn\, \frac{v_0 n\, -\, (v_0 n)^2 / 8}{(1\, -\, v_0 n / 4)^3}\,,
\end{equation}
and mean field $U_{\rm CS}(n,T)$ calculated with the help of Eq.~(\ref{eq:determ-u}) is
\begin{equation}
U_{\rm CS} (n,T)\ =\ -3T\, \left[1 - \frac{1-v_0 n/12}{(1 - v_0 n/4)^3} \right]\,.
\end{equation}

\begin{figure}
\begin{center}
\begin{minipage}{.65\textwidth}
\centering
\includegraphics[width=\textwidth]{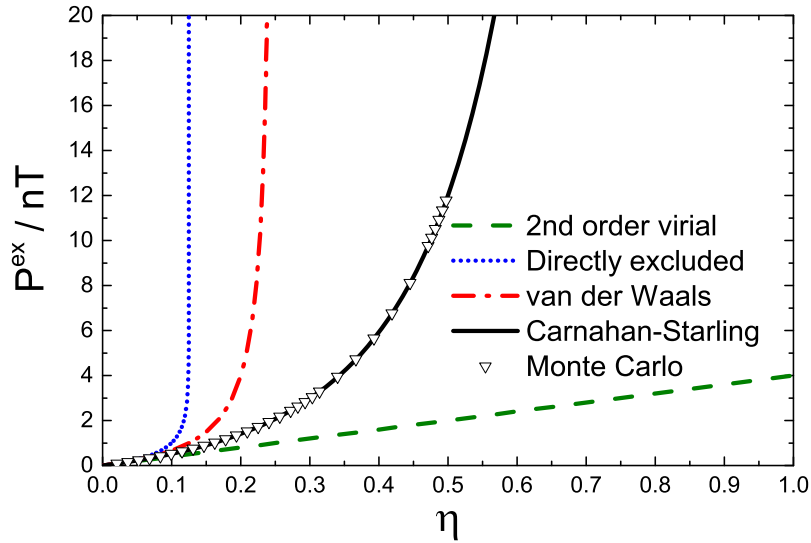}
\end{minipage}
\caption{(Color online)
The dependence of the quantity $P^{\rm ex} / nT$ on the packing fraction $\eta$
for four different models:
the 2nd order virial expansion (dashed green line),
the model of directly excluded-volume (dotted blue line),
the van der Waals equation of state (dash-dotted red line),
and the Carnahan-Starling equation of state (solid black line).
Monte Carlo data from Ref.~\cite{HSMC} is depicted by open triangles.
}
\label{fig:PexvsEta}
\end{center}
\end{figure}
In Fig.~\ref{fig:PexvsEta} we plot the dependence of the excess pressure
normalized by the ideal gas pressure, $P^{\rm ex} / nT$, on the packing
fraction, $\eta = v_0n/4$, for four different hard-sphere models:
the 2nd order virial expansion (dashed green line),
the model of directly excluded-volume (dotted blue line),
the van der Waals equation of state (dash-dotted red line), and
the Carnahan-Starling equation of state (solid black line).
Monte Carlo data from Ref.~\cite{HSMC} on the compressibility of hard spheres is
depicted by open triangles.
We note that all four models coincide up to 2nd order virial expansion.
We also note that $P^{\rm ex} / nT \equiv Z - 1$, where $Z$ is the
compressibility factor.
It is evident, see Fig.~\ref{fig:PexvsEta}, that the packing fraction is bounded from above
for these three approaches:
$\eta<0.125$ in the model of directly excluded-volume,
$\eta<0.25$ in the van der Waals model, and $\eta<1$ in the Carnahan-Starling model.
Note, the dense packing limit for hard spheres is $\eta = \pi / (3\sqrt{2}) \simeq 0.76$.
All the approaches coincide at small values of $\eta$ but start to diverge
significantly for $\eta>0.1$.
Comparison with Monte Carlo data indicates that the Carnahan-Starling equation
of state is the most accurate of the considered and can describe well Monte Carlo data
at least at up to $\eta=0.5$.

The presented mean-field approach allows one to conveniently implement even
more accurate equation of state for a gas of rigid spheres,
such as one from Ref.~\cite{ma1986equation} or another one given as the appropriate
Pad\'{e} approximation, for instance see \cite{anch}.
These approaches, however, are beyond the scope of the present study.

\subsection{Virial expansion for multi-component gas}
The virial expansion for classical multi-component gas of $f$ particle species
can be written as
\begin{equation}
p(\bvar n, T)\ =\ T \sum_{i=1}^f n_i + T \sum_{i,j} b_{ij}(T) n_i n_j
+ \sum_{i,j,k} b_{ijk}(T) n_i n_j n_k + \ldots\, =\,
T \sum_{i=1}^f n_i + P^{\rm ex} (\bvar n, T)\,.
\label{eq:virialmulti}
\end{equation}
Here $b_{ij},\ b_{ijk}, \ldots$ are the virial coefficients.
To get mean field $U_i(\bvar n, T)$, which corresponds to $i$-th species, we
express it as power series
\begin{equation}
U_i (\bvar n, T)\ =\ T \sum_j a_j^i(T) n_j\, +\, \sum_{j,k} a_{jk}^i(T) n_j n_k + \ldots \ .
\label{eq:Umvirial}
\end{equation}
Substituting $U_i$ in form \eqref{eq:Umvirial} into Eqs.~\eqref{eq:consistency-multi}
which are the conditions of thermodynamic consistency
we get a closed system of linear equations with respect to coefficients
$a_{j}^i,\, a_{jk}^,$, etc.
By solving this system we determine the mean fields $U_i$ in an unambiguous way from
the given virial expansion~\eqref{eq:virialmulti}.

As an example let us consider a second order virial expansion, i.e.
$P^{\rm ex} (\bvar n, T) = T \sum_{i,j} b_{ij}(T) n_i n_j$.
In this case  mean fields can be expressed as
$U_i (\bvar n, T) = T \sum_{j} a_{j}^i(T) n_j$.
Then, system of equations for coefficients $a_{j}^i$ can be solved explicitly
and the solution is
\begin{eqnarray}
a_j^i(T) & = & 2 b_{ij} (T),
\label{eq:solution-2nd-virial}
\\
U_i (\bvar n, T) & = & 2 T \sum_{j} b_{ij}(T) n_j.
\end{eqnarray}
Note, because of the evident symmetry of coefficients $b_{ij} = b_{ji}$ we obtain from
(\ref{eq:solution-2nd-virial}) the
symmetry of indexes of coefficients which determine mean field, $a_j^i = a_i^j$.
This statement is in full consistency with the mean-field property
(\ref{eq:symmetry}) and confirms it explicitly.

\subsection{Excluded-volume in multi-component gas: van der Waals
excluded-volume procedure I}
\label{sec:exclud-vol-1}
One way to extend the van der Waals excluded-volume procedure on multi-component
system of particles with different proper volumes $v_i$ is to replace the total
volume $V$ of ideal gas by the total available volume $V-\sum_i v_i N_i$.
This procedure was done in thermodynamically consistent way in
Ref.~\cite{Yen1997} by inserting the total available volume into the grand
partition function.
As a result, for a given point in $(T, \bs \mu)$-plane one obtains a transcendental
equation with respect to the system pressure $p(T, \bs \mu)$
\begin{equation}
p(T, \bs \mu)\, =\, \sum_{i=1}^f\, p_i^0 \big[T, \mu_i - v_i \, p(T, \bs \mu)\big]\,.
\label{eq:pressure-multi-excl}
\end{equation}
Here $p_i^0(T,\mu_i)$
is the ideal gas pressure
of species $i$.
In the Boltzmann limit, the expression for pressure \eqref{eq:pressure-multi-excl}
reads as
\begin{equation}
p(T, \bs \mu)\, =\, \sum_{i=1}^f \, p_i^0(T,\mu_i)\,
\exp{\left[ - \frac{v_i\,p(T, \bs \mu)}{T} \right]}\,.
\label{eq:pressure-multi-excl-a}
\end{equation}

Let us show how the mean-field approach can be used to conveniently
formulate a thermodynamically consistent procedure for multi-component gas
when we use the same prescription regarding the available volume.
To do that we substitute the excluded-volume $v_0\,n$ in expression
\eqref{eq:PVDW} for the single-component van der Waals excess pressure by the
total excluded-volume $\sum_i v_i n_i$ of multi-component gas.
We also treat $n$ as a total density of all particle species.
The excess pressure then reads
\begin{equation}
P_{\rm VdW}^{\rm ex} (\bvar n,T)\, =\, T \left( \sum_{j=1}^f n_j \right)
\frac{\sum_{i=1}^f v_i n_i}{1 - \sum_{i=1}^f v_i n_i} \, .
\label{eq:P-multi-excl}
\end{equation}
In order to perform calculations of thermodynamic variables one needs to
determing the mean fields $U_i$ by
solving the
system of equations \eqref{eq:consistency-multi}.
We remind that this system manifests the conditions of thermodynamic consistency.
In this particular case, solution for $U_i$ can be obtained in closed form
\begin{equation}
U_i(\bvar n, T)\, =\, T \, \frac{v_i \sum_{j=1}^f n_j}{1-\sum_{j=1}^f v_j n_j} - T \,
\ln{\left(1-\sum_{j=1}^f v_j n_j\right)} \,.
\label{eq:PU-multi-excl}
\end{equation}
With mean fields $U_i$ in hands we can now express the particle densities $n_i$,
which in the Boltzmann approximation read
\begin{equation}
n_i(T, \bs \mu)\, =\, \left(1-\sum_{j=1}^f v_j n_j\right) \, n_i^0(T, \mu_i) \,
\exp\left(-\frac{v_i \sum_{j=1}^f n_j}{1-\sum_{j=1}^f v_j n_j}\right)\,.
\label{eq:ni-multi-excl}
\end{equation}
We recall that the total pressure in the system can be written as
\begin{eqnarray}
p(T, \bs \mu) & = & T \left( \sum_{i=1}^f n_i \right)
+ P_{\rm VdW}^{\rm ex} [\bvar n(T, \bs \mu), T]\,
\nonumber \\
& = & T \frac{\sum_{i=1}^f n_i}{1 - \sum_{i=1}^f v_i n_i} \,.
\label{eq:pressure-multi-excl-2}
\end{eqnarray}
Substituting \eqref{eq:ni-multi-excl} into the numerator of \eqref{eq:pressure-multi-excl-2}
one can easily obtain Eq.~\eqref{eq:pressure-multi-excl-a}.
This excluded volume procedure has an advantage due to its simplicity:
just one transcendental equation \eqref{eq:pressure-multi-excl-a} needs to
be solved.
Meanwhile, this formulation is not consistent with a second order virial expansion
for a multi-component gas of hard spheres.

\subsection{Excluded-volume in multi-component gas: van der Waals
excluded-volume procedure II}
Let us now formulate an excluded-volume procedure for
multi-component gas which is consistent with second-order virial expansion.
Let $b_{ij}, i,j=1\ldots f$ denote the virial coefficients.
With account for different radii of particles $r_i$ and $r_j$ which belong to
different species $i$ and $j$, respectively, the virial coefficients
for hard-sphere model can be parameterized as
\begin{equation}
b_{ij}\, =\, \frac{2}{3} \pi (r_i + r_j)^3\,.
\end{equation}
The pressure in second order virial
expansion is then written as
\begin{equation}
p\ =\ T \sum_{i=1}^f n_i\, +\, T \sum_{i,j=1}^f b_{ij} n_i n_j\,.
\label{eq:P-multi-VdW-0}
\end{equation}

On the other hand, using a van der Waals prescription (\ref{eq:VDW}) one can
write the van der Waals pressure of multi-component
gas as a sum of partial pressures $p_i$ as
\begin{equation}
p\, =\, \sum_{i=1}^f\, p_i\, =\, \, \sum_{i=1}^f\, \frac{ T n_i}{1 - \sum_{j=1}^f \tilde{b}_{ij}\, n_j} \,,
\label{eq:p-multi-vdw-1}
\end{equation}
where $\tilde{b}_{ij}$ are the so far unknown coefficients.
In order for it to be consistent with the second order virial expansion
the van der Waals pressure (\ref{eq:p-multi-vdw-1}) should coincide with (\ref{eq:P-multi-VdW-0}).
up to second order expansion with respect to particle density.
Hence, it is necessary to constrain the coefficients $\tilde{b}_{ij}$ as:
$\tilde{b}_{ii}=b_{ii}$ and $\tilde{b}_{ij}+\tilde{b}_{ji}=2b_{ij}$.
The choice of the cross-terms $\tilde{b}_{ij}$ is not unique,
for instance, one may let $\tilde{b}_{ij} = b_{ij}$ or, as was proposed
in Ref.~\cite{Gorenstein1999}, 
$\tilde{b}_{ij} = 2 b_{ii} b_{ij}/(b_{ii}+b_{jj})$.

So, if we determine coefficients $\tilde{b}_{ij}$, then we determine
the van der Waals equation of state (\ref{eq:p-multi-vdw-1})
for the multi-component gas of different
species.
From this equation of state
one can obtain the corresponding excess pressure
\begin{equation}
P^{\rm ex} (\bvar n,T)\, =\, T \sum_{i=1}^f \left( \frac{n_i\,
\sum_{j=1}^f\tilde{b}_{ij} n_j}{1 - \sum_{j=1}^f \tilde{b}_{ij} n_j} \right)\,.
\label{eq:P-multi-VdW}
\end{equation}
Using this expression
in equations \eqref{eq:consistency-multi}, which guarantee  thermodynamic
consistency, one can also obtain mean fields $U_i$ in a closed form
\begin{equation}
U_i(\bvar n, T)\,
=\, T \, \sum_{j=1}^f \frac{\tilde{b}_{ij} n_j}{1-\sum_{k=1}^f \tilde{b}_{jk} n_k}\,
-\, T \, \ln\left(1-\sum_{j=1}^f \tilde{b}_{ij} n_j\right).
\label{eq:U-multi-VdW}
\end{equation}
Then, exploiting the mean fields $U_i$ from (\ref{eq:U-multi-VdW}),
the particle densities $n_i$ in the Boltzmann approximation are obtained in the
following form
\begin{equation}
n_i(T, \bs \mu)\, =\, \left(1-\sum_{j=1}^f \tilde{b}_{ij}\, n_j\right) \, n_i^0(T, \mu_i) \,
\exp\left(-\sum_{j=1}^f \frac{\tilde{b}_{ij}\, n_j}{1-\sum_{k=1}^f \tilde{b}_{jk}\, n_k}\right),
\quad i=1\ldots f.
\label{eq:ni-multi-VdW}
\end{equation}
Expression \eqref{eq:ni-multi-VdW} represents a system of transcendental equations
for particle densities $n_i$ which needs to be solved numerically for every given
point in $(T, \bs \mu)$-plane.
The total pressure can be expressed as a sum
\begin{eqnarray}
p(T, \bs \mu)\,  =\,  \sum_{i=1}^f \, p_i(T, \bs \mu)\,,
\qquad {\rm where} \qquad
p_i(T, \bs \mu)\,  =\,  \frac{T n_i}{1 - \sum_{j=1}^f \tilde{b}_{ij}\, n_j} \,.
\label{eq:pi-multi-VdW-1}
\end{eqnarray}
If for a given $(T, \bs \mu)$-point the system of transcendental equations
\eqref{eq:ni-multi-VdW} is solved, then substituting $n_i$ into numerator in (\ref{eq:pi-multi-VdW-1})
and rewriting the argument of exponent with the help of $p_i$ from
(\ref{eq:pi-multi-VdW-1}) in the same way as in (\ref{eq:pressure-multi-excl-2})
one can easily show that the ``partial'' pressures $p_i(T, \bs \mu)$ from
(\ref{eq:pi-multi-VdW-1}) are solutions of the following system of
transcendental equations
\begin{equation}
p_i(T, \bs \mu)\, =\, p_i^0(T, \mu_i) \,
\exp{\left[-\, \frac{\sum_{j=1}^f \tilde{b}_{ij}\, p_j(T, \bs \mu)}{T} \right]} \,.
\label{eq:pi-multi-VdW}
\end{equation}
The same system of equations was obtained in Ref.~\cite{Gorenstein1999} as well
by direct calculation of the grand partition function using excluded-volume
procedure for multi-component gas.

The presented model is fully consistent with the 2nd order virial expansion,
however, to get the solution
one needs to  solve one of the two equivalent systems of transcendental equations:
either system \eqref{eq:ni-multi-VdW} for particle densities
or system \eqref{eq:pi-multi-VdW} for ``partial'' pressures.
On the other hand, the approach formulated in previous subsection,
while not fully consistent with the 2nd order virial expansion,
is much simpler as only a single transcendental equation needs to be solved,
and, thus, this approach may be more practical to use.
%

%
\section{Mean field approach in hadron-resonance gas model}
\label{sec6}
Statistical models are very successful in description of wide range of data on
mean hadron multiplicities in various heavy-ion collision experiments.
The most well-known
is the ideal hadron-resonance gas (I-HRG) model formulated in the framework of
grand canonical ensemble.
The I-HRG model represents a thermodynamic
statistical system of non-interacting hadrons and resonances.
The pressure in such system is given by
\begin{equation}
p\, =\, \sum_{i=1}^f p_i(T,\mu_i)\, =\, \sum_{i=1}^f g_i \int \frac{d^3k}{(2\pi)^3}
\frac{\bvar k^2}{\sqrt{m_i^2 + \bvar k^2}}
\, \left[\exp\left(\frac{\sqrt{m_i^2 + \bvar k^2} - \mu_i}{T}\right)+a_i\right]^{-1}\,,
\label{eq:pIHRG}
\end{equation}
where $g_i$ is the degeneracy of hadron species $i$, $T$ is the temperature,
$a_i$ equals $-1$ for bosons, $+1$ for fermions and it equals $0$ for
the Boltzmann gas.
The chemical potential $\mu_i$ consists of contributions from baryon,
strangeness and electric charge chemical potentials:
\begin{equation}
\mu_i = B_i \mu_B + S_i \mu_S + Q_i \mu_Q.
\end{equation}
The particle density of hadron species $i$ reads as
\begin{equation}
n_i (T, \mu_i)\, =\, g_i \int \frac{d^3k}{(2\pi)^3}
\, \left[\exp\left(\frac{\sqrt{m_i^2 + \bs k^2} - \mu_i}{T}\right)+a_i\right]^{-1} \,.
\label{eq:niIHRG}
\end{equation}
Considering $T$ and $\mu_B$ as free parameters and properly taking into account
the decays of unstable hadrons one can fit multiplicity ratios measured in various
heavy-ion collision experiments.
Chemical potentials $\mu_S$ and $\mu_Q$ are usually expressed as functions of
$T$ and $\mu_B$ by fixing the total net strangeness $\langle S \rangle = 0$ and
electric to baryon charge ratio
$\langle Q \rangle / \langle B \rangle = Z/A \approx 0.4$.

In the interacting hadron resonance gas formulated in the mean-field approach
(MF-HRG) expressions \eqref{eq:pIHRG} and \eqref{eq:niIHRG} for pressure and
particle density now read
\begin{eqnarray}
\label{eq:niMFHRG-1}
p & = & \sum_{i=1}^f \frac{g_i}{6\pi^2}\int_0^{\infty} \frac{k^4\,dk}{\sqrt{m_i^2 + k^2}}
\, f_i(k,m_i;T,\mu_i,\bvar n)\, +\, P^{\rm ex} (\bvar n, T)\,,
\\
n_i & = & \frac{g_i}{2\pi^2}\int_0^{\infty} k^2\,dk \, f_i(k,m_i;T,\mu_i,\bvar n)\,,
\label{eq:niMFHRG}
\end{eqnarray}
where
\begin{eqnarray}
f_i(k,m_i;T,\mu_i,\bvar n)\ =\
\left\{\exp\left[\frac{\sqrt{m_i^2 + k^2}-\mu_i + U_i (\bvar n, T)}{T}\right]+a_i\right\}^{-1}\,,
\label{eq:niMFHRG-2}
\end{eqnarray}
is the distribution function of the $i$th species, in which the free dispersion
law is shifted by the mean field $U_i$.

In our calculations we include all non-strange and strange hadrons that are
listed in Particle Data Tables~\cite{pdg}.
This includes mesons up to $f_2(2340)$ and (anti)baryons up to $N(2600)$.
The finite width of resonances and their Breit-Wigner form is taken into account
in a similar way as in THERMUS package~\cite{THERMUS}, by making the following
modification of the integration of distribution function:
\begin{equation}
\int d^3k \, f_i(k,m_i;\bvar n, T, \mu_i)\, \to \, \int d m \int d^3k \,
\rho_i(m) \, f_i(k,m;\bvar n, T, \mu_i) \,,
\label{eq:niMFHRG-3}
\end{equation}
where $\rho_i(m)$ is the properly normalized mass distribution for
hadron $i$. We note that for stable hadrons $\rho_i(m) = \delta(m-m_i)$,
while for resonances it has a Breit-Wigner form.

In order to perform calculations one needs to solve self-consistently a system
of transcendental equations for hadron densities \eqref{eq:niMFHRG}.
In the present study we include effects of excluded-volume in the framework of
the mean-field approach and we consider only the case when the repulsive
interaction between all kinds of hadrons is the same, i.e. all hadrons have the
same hard-core radius $r$.
In this case all mean fields coincide ($U_i \equiv U$) and depend just on the
total density of all hadrons, $n = \sum_i n_i$.
Therefore, only single transcendental equation for total density
$n$ needs to be solved. This equation is solved numerically using the secant method.

We consider three different excluded-volume parameterizations of mean field:

a) The model of directly excluded-volume:
\begin{eqnarray}
P^{\rm ex} (n,T) & = & -\, \frac{T}{2v_0} \ln{\left(1 - 2v_0 n\right)} - nT\,,
\\
U(n,T) & = & -\, T \ln{\left(1 - 2 v_0 n\right)}\,,
\end{eqnarray}
where $v_0$ is four times the intrinsic volume of hadron.

b) The van der Waals excluded-volume procedure:
\begin{eqnarray}
P^{\rm ex} (n,T) & = & Tn \frac{v_0 n}{1-v_0 n},
\\
U(n,T) & = & T \frac{v_0 n}{1 - v_0 n} - T\ln{\left(1 - 2 v_0 n\right)} \,.
\end{eqnarray}

c) The Carnahan-Starling model:
\begin{eqnarray}
P^{\rm ex} (n,T) & = & Tn\, \frac{v_0 n\left(1 - \frac 18 v_0 n \right) }{\left(1 - \frac 14 v_0 n \right)^3}\,,
\\
U(n,T) & = & -3T\, \left[1 - \frac{1 - \frac{1}{12} v_0 n}{\left(1 - \frac 14 v_0 n \right)^3} \right]\,.
\end{eqnarray}

Note, all three cases are consistent with second-order virial expansion
for a gas of hard spheres.

We would like to study properties of a hadron-resonance gas
at the stage of a chemical freeze-out during heavy-ion collisions.
As a result of analysis of particle multiplicities at various heavy-ion
collision experiments the collision energy dependence of temperature and
baryochemical potential at the chemical freeze-out can be parametrized
as~\cite{Cleymans2006}
\begin{eqnarray}
T(\sqrt{s_{_{\rm NN}}}) & = & 0.166\,{\rm GeV} - 0.139\,{\rm GeV}^{-1} \mu_B^2
-  0.053\,{\rm GeV}^{-3} \mu_B^4, \label{eq:Tfrz} \\
\mu_B(\sqrt{s_{_{\rm NN}}}) & = & \frac{1.308\,{\rm GeV}}{1 + 0.273\,{\rm GeV}^{-1} \sqrt{s_{_{\rm NN}}}}. \label{eq:muBfrz}
\end{eqnarray}
At the highest collision energies this parametrization
yields $T \approx 166$~MeV and $\mu_B \approx 0$.
We note that recent lattice QCD calculations~\cite{lattice-1,lattice-2}
indicate the QCD crossover transition temperature of $T\sim155$~MeV, a lower value than given by \eqref{eq:Tfrz}. Recent analysis of particle yields and ratios, as well as fluctuations of conserved charges, indeed indicate a lower chemical freeze-out temperature~\cite{Stachel2013,Floris,PBM2014}. Nevertheless, these details play very minor role in the context of calculations in the present work. Therefore, we will use parametrization given by Eqs.~\eqref{eq:Tfrz} and \eqref{eq:muBfrz} to study collision energy dependence of chemical freeze-out properties.

\begin{figure}
\begin{center}
\begin{minipage}{.65\textwidth}
\centering
\includegraphics[width=\textwidth]{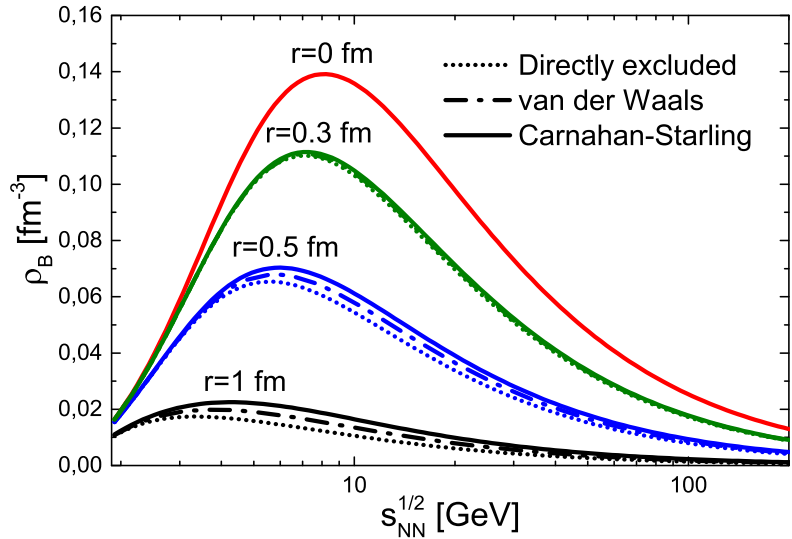}
\end{minipage}
\caption{(Color online) Energy dependence of the net-baryon density calculated
in the mean field approach for different values of hadron hard-core radii using
three different excluded-volume procedures:
the model of directly excluded-volume (dotted line),
the van der Waals equation of state (dash-dotted line),
and the Carnahan-Starling equation of state (solid line).}
\label{fig:rhoB}
\end{center}
\end{figure}

In order to account for incomplete equilibration in the strangeness sector a
suppression factor $\gamma_S$ is usually introduced in the distribution function
and its energy dependence parametrization reads as~\cite{Becattini2006}
\begin{equation}
\gamma_S(\sqrt{s_{_{\rm NN}}}) = 1 - 0.396 \exp\left(-1.23 \frac{T}{\mu_B}\right).
\end{equation}
This factor is introduced into $f_i(k,m;T,\mu_i,\bvar n)$ in expression for
density and pressure as
\begin{equation}
f_i(k,m;T,\mu_i,\bvar n)\, =\, \left[\gamma_S^{-|s_i|}\,
\exp\left(\frac{\sqrt{k^2+m_i^2}-\mu_i + U_i (\bvar n, T)}{T}\right)+a_i\right]^{-1}\,,
\end{equation}
where $|s_i|$ is the absolute strangeness content for hadron species $i$,
i.e. $|s_i|$ is the total number of strange quarks and anti-quarks in $i$-th hadron.
The results of the calculations for energy dependence of the net-baryon density
on the center-of-mass collision energy, $\sqrt{s_{_{\rm NN}}}$, is presented in
Fig.~\ref{fig:rhoB}.
We note that results for the van der Waals excluded-volume model are consistent
with those in Ref.~\cite{Begun2013} obtained with the help
of a THERMUS package.
The inclusion of a hard-core repulsion results in a shift of a maximum in
net-baryon density energy dependence to the region of lower collision energies
and it also broadens the peak.
For instance, for a hard-core radius of 0.5~fm the maximum is shifted from $\sqrt{s_{_{\rm NN}}} \simeq 8$~GeV to $\sqrt{s_{_{\rm NN}}} \simeq 6$~GeV. These energies of maximum net-baryon density at freeze-out lie firmly in the range of a future Compressed Baryonic Matter (CBM) experiment at the Facility for Antiproton and Ion Research (FAIR).

\begin{figure}
\begin{center}
\begin{minipage}{.49\textwidth}
\centering
\includegraphics[width=\textwidth]{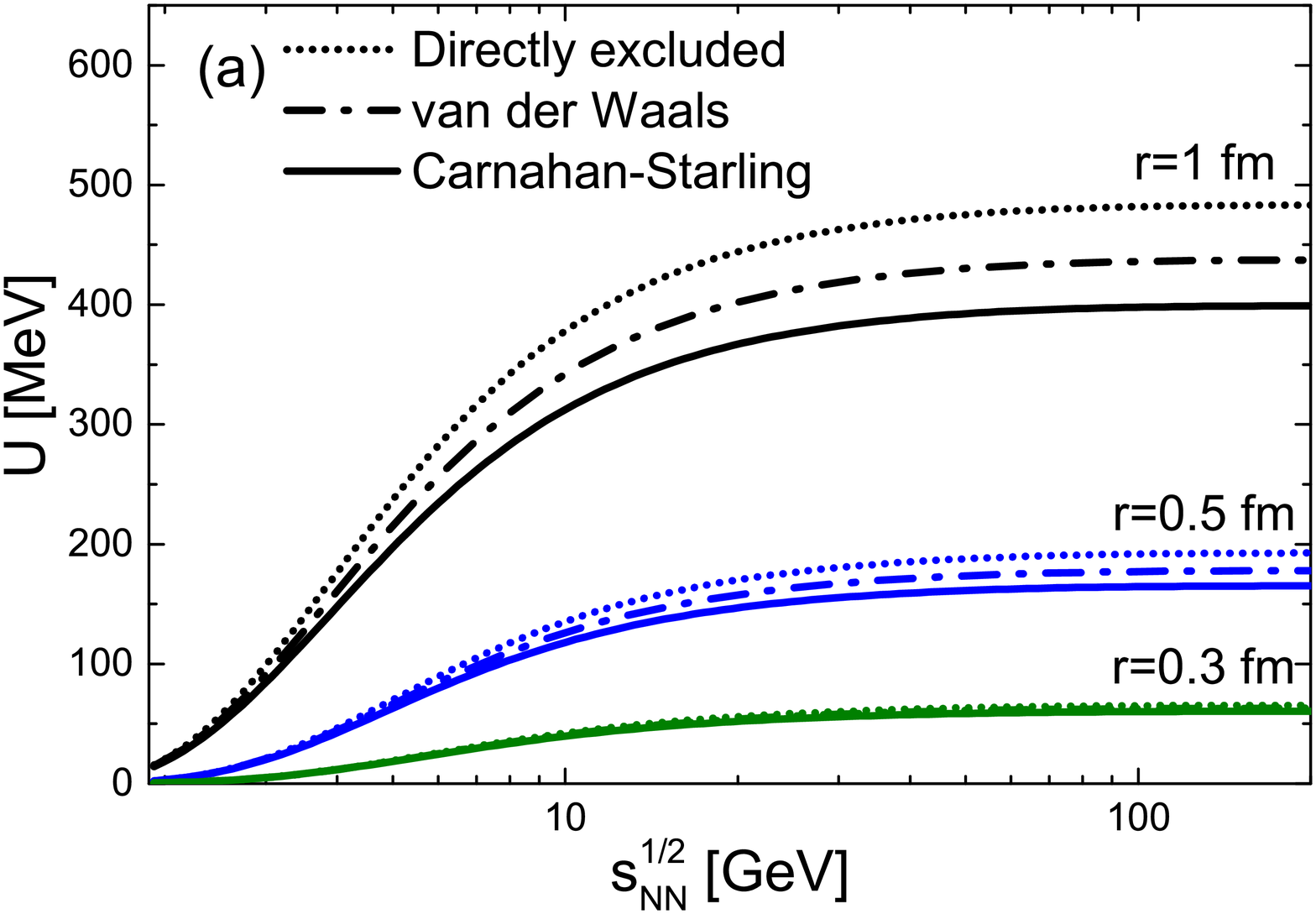}
\end{minipage}
\begin{minipage}{.49\textwidth}
\centering
\includegraphics[width=\textwidth]{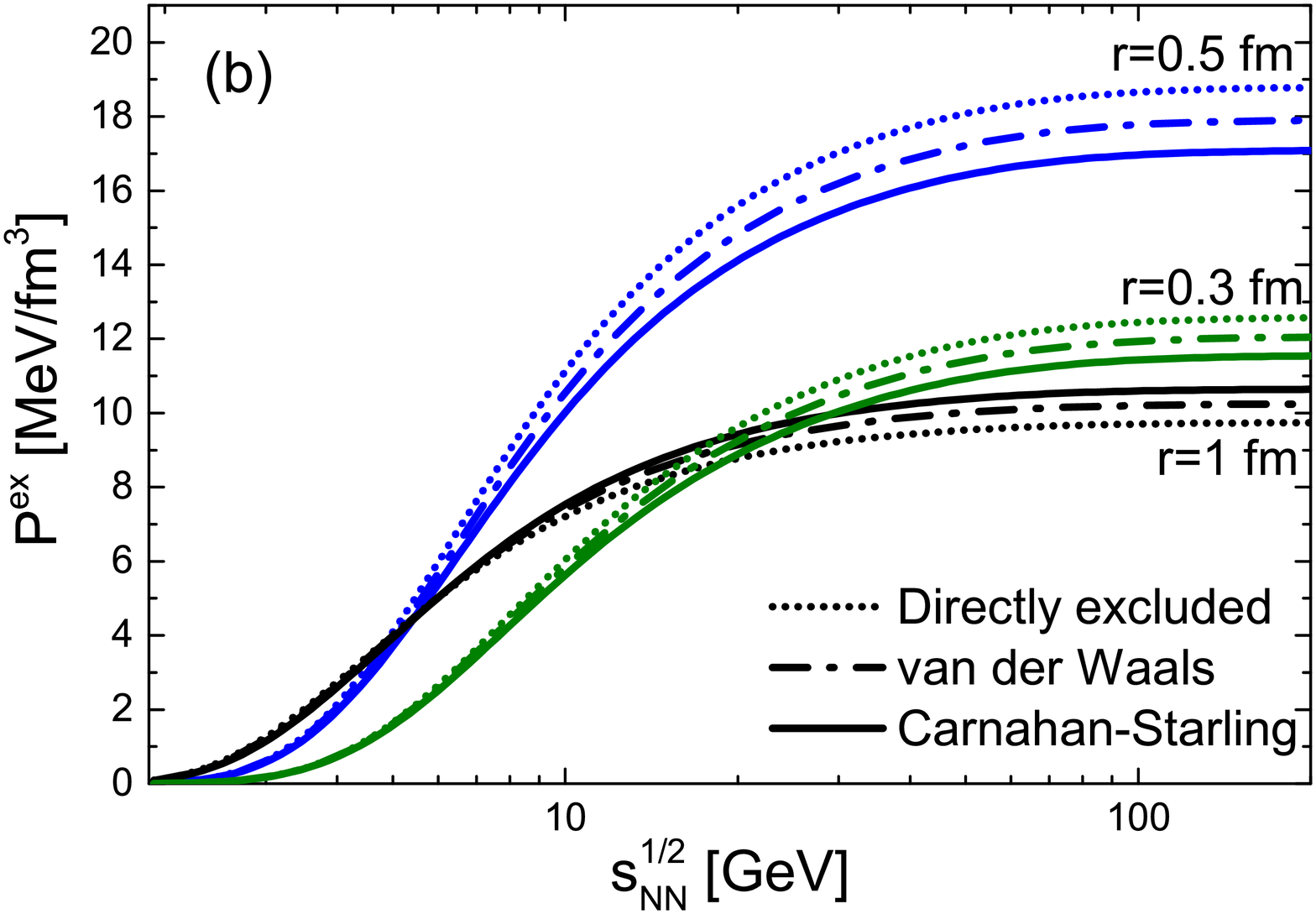}
\end{minipage}
\caption{(Color online) Energy dependence of (a) mean field $U$ and
(b) excess pressure $P^{\rm ex}$ calculated for different hadron hard core radii
in the mean field approach using three different excluded-volume procedures:
the model of directly excluded-volume (dotted line),
the van der Waals equation of state (dash-dotted line),
and the Carnahan-Starling equation of state (solid line).
}
\label{fig:UandP}
\end{center}
\end{figure}

The energy dependence of the mean field and excess pressure is depicted
in Fig.~\ref{fig:UandP}.
It is seen that they both increase with collision energy, and the difference
between the considered excluded-volume procedures increases with collision
energy.
Meanwhile, one can see the saturation of the mean field in the region
$\sqrt{s_{_{\rm NN}}} \ge 20A$~GeV for all of the excluded-volume models considered in
the present paper.
It is especially pronounced for the smaller hadron radii, e.g. at $r \le 0.5$~fm.
Hence, we can conclude that the effects of hard-core repulsion
remain approximately constant at collisions energies which are bigger
than $20A$~GeV.

One can see from Fig.~\ref{fig:UandP} (right) that the excess pressure
is a non-monotonous function of the hadron hard-core radius at high collision energies.
To explain this we recall that, by definition, $P^{\rm ex}(n,T)$
is the difference between the interacting system pressure and the ideal gas pressure at the
same values of temperature and particle density, e.g., $P^{\rm ex}(n,T) = p - nT$
in case of Boltzmann statistics.
In excluded-volume models this quantity is proportional to the total density $n$
and it also depends on the packing fraction, $\eta \equiv v_0 n/4$.
On the other hand, at fixed $T$ and $\mu$, the presence of excluded
volume leads to the reduction of particle density $n(T,\mu)$ compared to the
ideal gas density $n^{\rm id}(T,\mu)$, as seen from Eq.~\eqref{eq:particle-dens-3},
and, generally, $n(T,\mu)$ decreases with increase of hard-core radius $r$.
Naturally, $n(T,\mu) \to 0$ when particle self-volume $v_0 \to \infty$.
For this reason $P^{\rm ex}$, which is proportional to $n$, goes to zero as
$r \to \infty$.
On the other hand, also $P^{\rm ex}=0$ when $r=0$ since there are no interactions when $r=0$.
This entails that, at fixed $T$ and $\mu$, the dependence of $P^{\rm ex}$ on
hard-core radius $r$ is non-monotonic and has a maximum.

\begin{figure}
\begin{center}
\begin{minipage}{.65\textwidth}
\centering
\includegraphics[width=\textwidth]{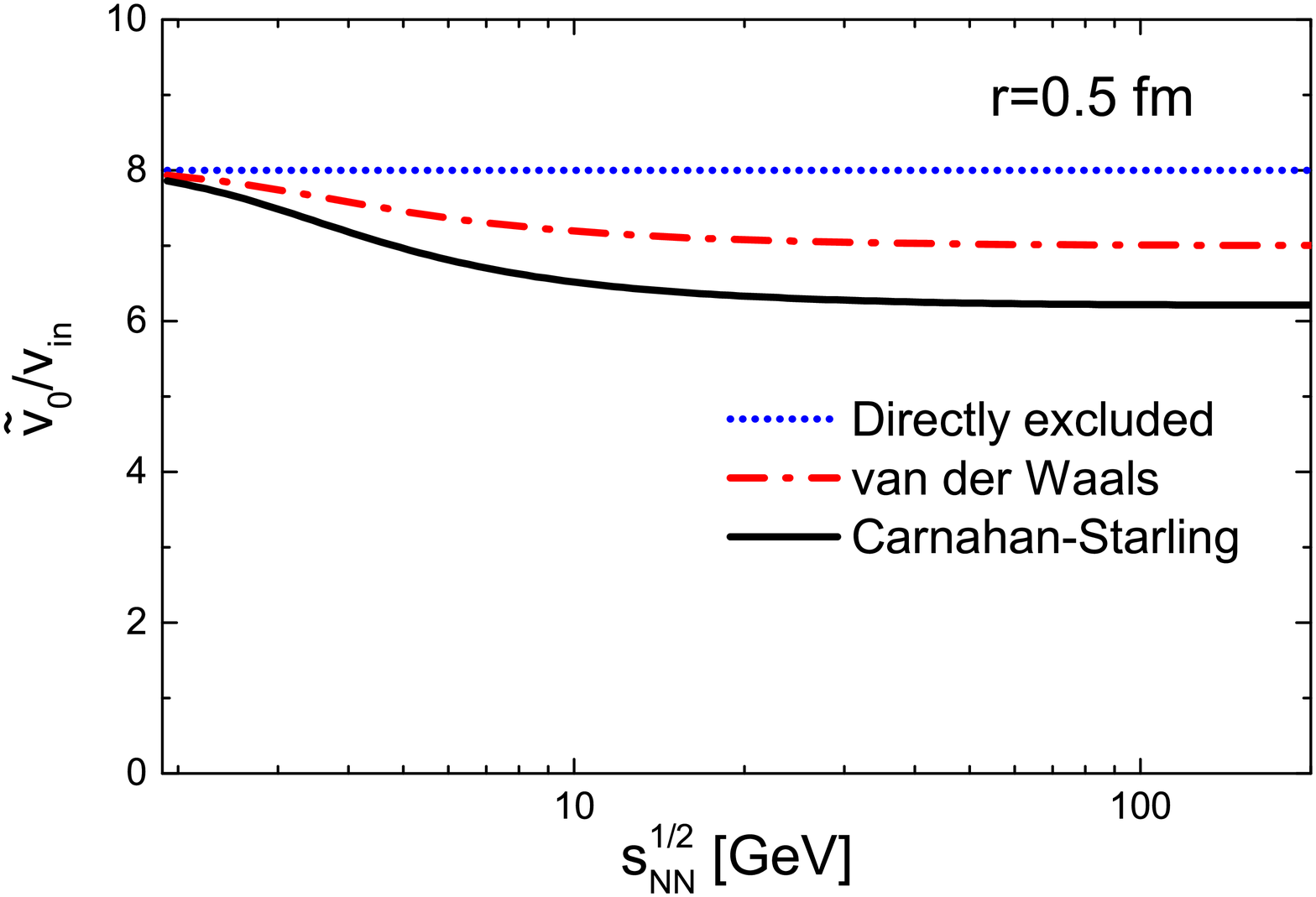}
\end{minipage}
\caption{(Color online) Energy dependence of the effective single particle
self-volume calculated in accordance with Eq.~\eqref{eq:v0-2} using three
different excluded-volume procedures:
the model of directly excluded-volume (dotted blue line),
the van der Waals equation of state (dash-dotted red line),
and the Carnahan-Starling equation of state (solid black line).
Intrinsic particle volume $v_{\rm in} = 4 \pi r^3 / 3$, where we take $r = 0.5$~fm.
}
\label{fig:effV}
\end{center}
\end{figure}

The effective single particle proper volume $\tilde{v}_0$ generated by the
repulsive mean field can be calculated using
Eq.~\eqref{eq:v0-2} and is plotted in Fig.~\ref{fig:effV}.
We normalize the value of $\tilde{v}_0$ by the intrinsic particle volume
$v_{\rm in} = 4 \pi r^3 / 3$, where we use $r = 0.5$~fm.
It is seen that for the model of directly excluded-volume the value of
$\tilde{v}_0$ is fixed at $\tilde{v}_0 = 8 v_{\rm in}$, which can be interpreted
as the volume around the center of a spherical particle, inside of which no
other center of a spherical particle can be located if the spheres are non-penetrable.
For the more realistic hard sphere equations of state
such as the van der Waals and Carnahan-Starling we have that $\tilde{v}_0 < 8 v_{\rm in}$.

\begin{figure}
\begin{center}
\begin{minipage}{.65\textwidth}
\centering
\includegraphics[width=\textwidth]{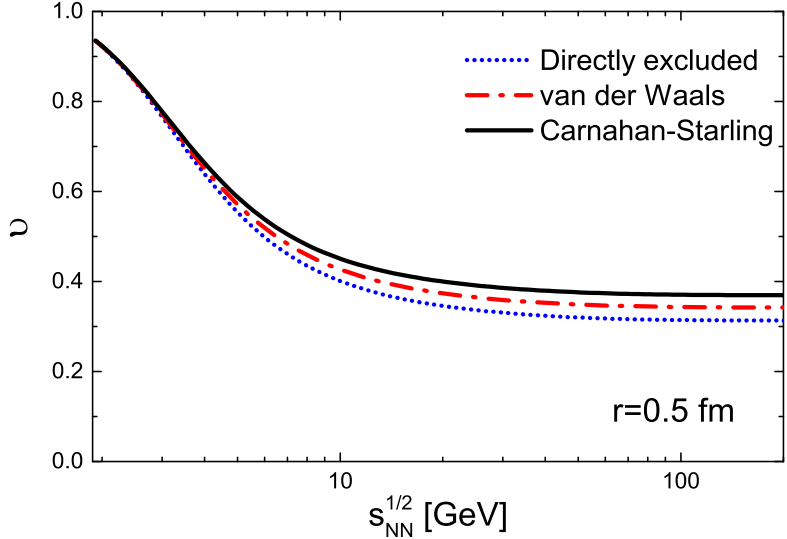}
\end{minipage}
\caption{(Color online) Energy dependence of of the factor $\vartheta$
which quantifies the decrease of the total available volume $\widetilde{V}$,
i.e. $\vartheta(n) = \widetilde{V}/V$, as well as
the suppression of quantum statistics due to repulsive mean field.
Calculations are done using three different excluded-volume procedures
and with the value of hadron hard core radius equal to 0.5~fm.}
\label{fig:suppr}
\end{center}
\end{figure}

As was discussed in Sec.~\ref{sec5}, the factor $\vartheta(n) = e^{-U(n,T)/T}$
quantifies the decrease of available volume $\widetilde{V}$, i.e.
$\vartheta(n) = \widetilde{V}/V$,
and suppression
of quantum statistics compared to
the case of ideal gas (see (\ref{eq:f0-quant}) and discussion after it).
We plot the energy dependence of this quantity for the hard-core radius of
0.5~fm in Fig.~\ref{fig:suppr}.
It is seen that, even at a moderate value of $r=0.5$~fm, the densities and also
the quantum statistics become significantly suppressed, especially at higher
collisions energies.
These results further indicate that, in most cases, appliance of the Boltzmann
approximation is quite sufficient for the studies of chemical freeze-out in
heavy-ion collisions.

\begin{figure}
\begin{center}
\begin{minipage}{.65\textwidth}
\centering
\includegraphics[width=\textwidth]{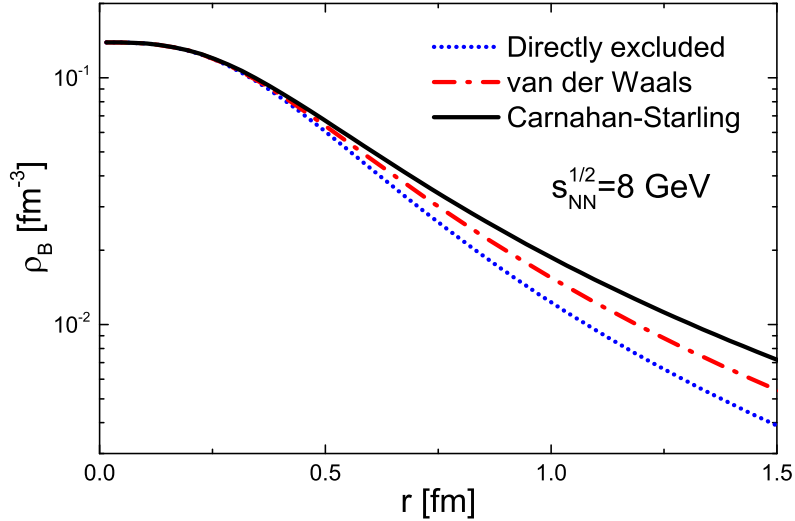}
\end{minipage}
\caption{(Color online) Dependence of the net-baryon density
on the value of the hard-core hadron radius.
Calculations are done for the values: $T=141$~MeV, $\mu_B = 411$~MeV and
$\gamma_S = 0.74$, which
correspond to the collision energy of $\sqrt{s_{_{NN}}} = 8$~GeV.
}
\label{fig:rhoBvsR}
\end{center}
\end{figure}

We would like to investigate the effects of the size of the hard-core hadron
radii $r$ that we use in our study.
In Fig.~\ref{fig:rhoBvsR} we plot the dependence of the net-baryon density on
the value of $r$ for a fixed collision energy of
$\sqrt{s_{_{\rm NN}}} = 8$~GeV, which corresponds to the following values of
parameters: $T=141$~MeV, $\mu_B=411$~MeV and $\gamma_S=0.74$.
It is seen that all three considered procedures give very similar results
for moderate values of hadron radii ($r\leq0.5$~fm), meanwhile
the relative difference for the net-baryon density given by different
excluded-volume approaches increases with value of $r$ and reaches about
$20-25\%$ at $r \gtrsim 0.9$~fm.
The situation is essentially the same for RHIC and LHC energies as well.
These results show that the choice of excluded-volume procedure becomes
important in the studies of chemical freeze-out if one considers large
values of hard-core hadron radii and in this case the van der Waals
procedure may become quite inaccurate.
Hence, if we are going to interpret our system as a gas of hard spheres then,
evidently, the Carnahan-Starling model gives the most accurate results among
the presented excluded-volume approaches.
However, in order to describe the lattice QCD data at $T<155$~MeV within
the EV-HRG it is necessary that the hard-core radius of hadrons is constrained
from above, namely $r \lesssim 0.5$~fm~(see, e.g., Ref.~\cite{VAG2015}).
For these values of $r$ the differences are no more than a few percent, and the standard van der Waals excluded-volume procedure appears to be
sufficient to describe the densities at the chemical freeze-out at all collision energies,
as seen in Figs.~\ref{fig:rhoB} and \ref{fig:rhoBvsR}.
We note that the particle number fluctuations in HRG may be more sensitive to
the excluded-volume effects \cite{Nikolajenko}, however, we do
not study this subject in the present work.

\section{Summary and conclusions}
\label{sec7}

In the presented thermodynamic mean-field approach we express interactions
in a multi-component gas in terms of mean fields, $U_i(n,T)$, and the excess
pressure, $P^{\rm ex} (n,T)$.
This approach provides a prescription to conveniently generalize known
phenomenological models of equation-of-state to the case of a system with
variable number of particles in the grand canonical formulation, which is of
vital significance for relativistic systems.

In the present work we formulated three different excluded-volume procedures
for the mixture of particles with the same hard-core radii in terms of
mean-fields: the directly excluded-volume, the van der Waals equation-of-state,
and the Carnahan-Starling equation-of-state.
The mean fields in excluded-volume models are linearly dependent on temperature
and we show that such kind of fields generate an effective self-volume of a particle.
As a result, the expressions for the total number of particles and the total energy
are reduced to the case of ideal gas, but with total volume $V$
suppressed by a density-dependent factor of $\vartheta(n) = \exp[-U(n,T)/T]$.
It was shown that the presence of excluded-volume does not change
average energy per particle in the Boltzmann limit.
In case of quantum statistics, in addition to volume suppression, the presence
of excluded volume leads also to suppression of the quantum statistical effects,
see Eqs.~(\ref{eq:energy-dens-7}), (\ref{eq:energy-dens-8}) and \eqref{eq:f0-quant}.

The three different excluded-volume procedures are then used in the
hadron-resonance gas model for studies of the collision energy dependence of
various parameters at chemical freeze-out in heavy-ion collisions.
In all three cases, the inclusion of hard-core repulsion results in a shift of
the net-baryon maximum location to lower collision energies, a result also
reported earlier~\cite{Begun2013}.
It is interesting to note that, for any excluded-volume model considered in the
present paper, there is a saturation of the mean field starting from collision energies
$\sqrt{s_{_{\rm NN}}} \simeq 20$~GeV, see Fig.~\ref{fig:UandP}.
As it is seen, the saturation is especially pronounced for hadron radii $r \le 0.5$~fm.
Hence, for collision energies $\sqrt{s_{_{\rm NN}}} \ge 20$~GeV we can take the
mean field as a constant one.
This saturation of the mean field also results in saturation of the effective
particle volume $\widetilde{v}_0 = (V - \widetilde{V})/N$ (Fig.~\ref{fig:effV}),
and in saturation of the available volume quantified by
the $\vartheta$-factor, $\vartheta(n) = \widetilde{V}/V$, (Fig.~\ref{fig:suppr}).
The latter means that suppression of quantum statistical effects which is
determined by the $\vartheta$-factor will be at the same level for all
collision energies $\sqrt{s_{_{\rm NN}}} \ge 20$~GeV.

The calculations also show that, for all collision energies, the differences between the three procedures
stay rather small at moderate values of hadron radii ($r\lesssim0.5$~fm),
the values also suggested by comparison of EV-HRG with lattice QCD.
This confirms the validity of the van der Waals excluded-volume procedure,
which is commonly used to include the effects of short-range repulsion in HRG.
The differences between considered models may only become significant at large and rather unrealistic
values of hadron radii ($r\gtrsim0.9$~fm), where the relative differences
for net-baryon density may reach about $20-25\%$.
This indicates that a choice of the model for repulsive interactions may
be significant when one considers hadrons of large size,
or when one deals with highly dense systems such as compact stars.
If one interprets hadrons as hard spheres, then the Carnahan-Starling procedure
is the most accurate of the three considered in our work.
The effects of relativistic contraction would inevitably distort the picture of
rigid spheres, however, this effect is expected to be small for a hadron gas.
Indeed, since the temperature of a hadron gas is significantly smaller than masses of all
hadrons except pions, than this effect may be significant only for thermal pions.
Since the majority of pions come from resonance decays, which again have a large
masses compared to temperature, then this effect can be expected to stay small even
for pions in general.

The presented approach allows one to consider as well a multi-component system
of particles of different sizes.
We use the mean-field approach to formulate two different
van-der-Waals-like excluded-volume procedures for this case. In the Boltzmann
limit, these approaches are shown to be consistent with previously proposed
formulations.
In both of these models, for any given values of temperature and chemical
potentials, one obtains a system of transcendental equations for densities of
all considered particle species, and this system needs to be solved numerically.

\begin{acknowledgments}
%
We would like to thank M.I. Gorenstein for useful discussions.
Publication is based on
the research provided by the grant support of the State Fund for Fundamental
Research (project No. F58/175-2014). The work was partially supported by the
LOEWE program of HIC for FAIR (Frankfurt am Main) and by the Program of
Fundamental Research of the Department of Physics and Astronomy of
National Academy of Sciences of Ukraine (project No. 0112U000056).

\end{acknowledgments}


\end{document}